\documentclass[usenatbib]{mn2e}
\usepackage{float}
\usepackage{graphicx}
\usepackage{amsmath}
\usepackage{placeins}
\usepackage{threeparttable}
\usepackage{mathptmx}
\usepackage{txfonts}
\usepackage[T1]{fontenc}
\usepackage{ae,aecompl}
\usepackage{pdfpages}
\usepackage{etex}
\usepackage{tablefootnote}
\usepackage{scalerel}
\usepackage{longtable}
\usepackage{hyperref}
\usepackage{tabularx}
\usepackage{threeparttable} 
\usepackage{xcolor} 

\hypersetup{colorlinks, citecolor=blue}
\newcommand\HI{H\protect\scaleto{$I$}{1.2ex}}

\newcommand{\kms}{\mbox{km\,s$^{-1}$}}
\begin{document}
\title[\HI{} 21cm observations and dynamical models of FGC 2366 ]
{\HI{} 21cm observations and dynamical modelling of the thinnest galaxy: FGC 2366}

\author[Aditya et al.]
       {K. Aditya$^{1}$ \thanks{E-mail: kaditya@students.iisertirupati.ac.in}, 
       Arunima Banerjee$^{1}$ \thanks{E-mail: arunima@iisertirupati.ac.in},
       Peter Kamphuis$^{2}$,
       Aleksandr Mosenkov$^{3,4}$\newauthor,
       Dmitry Makarov $^{5}$ 
       Sviatoslav Borisov$^{5,6,7}$\\
$^1$   Department of Physics, Indian Institute of Science Education and Research (IISER) Tirupati, Tirupati - 517507, India\\
$^2$   Ruhr Universität Bochum, Astronomisches Institut, Universitätsstrasse 150, D-44801 Bochum, Germany\\
$^3$   Department of Physics and Astronomy, N283 ESC, Brigham Young University, Provo, UT 84602, USA\\
$^4$   Pulkovo Observatory, Russian Academy of Sciences, St.\ Petersburg, 196140 Russia\\
$^5$   Special Astrophysical Observatory, Russian Academy of Sciences, Nizhnii Arkhyz, 369167 Russia\\
$^6$   Sternberg Astronomical Institute, M.V. Lomonosov Moscow State University, Universitetsky prospect 13, Moscow, 119234 Russia\\
$^7$   Department of Astronomy, University of Geneva, Chemin Pegasi 51, 1290 Versoix, Switzerland}
\maketitle

\begin{abstract}
Superthin galaxies are bulgeless low surface brightness galaxies with unusually high major-to-minor axes ratio of the stellar disc, i.e.,$10<a/b<20$. We present Giant Metrewave Radio Telescope (GMRT) \HI{} 21cm radio-synthesis observations of FGC 2366, the thinnest galaxy known with $a/b=21.6$. Employing the 3-D tilted-ring modelling using Fully Automated TiRiFiC (FAT), we determine the structure and kinematics of the \HI{}  gas disc, obtaining an asymptotic rotational velocity equal to 100 \kms and a total \HI{} mass equal to 10$^9 M_{\odot}$. Using $z$-band stellar photometry, we obtain a central surface brightness of 22.8 mag ${\rm{arcsec}}^{-2}$, a disc scale length of 2.6 kpc, and a scaleheight of 260 pc. Next, we determine the dark matter density profile by constructing a mass model and find that an NFW dark matter halo best fits the steeply-rising rotation curve. With the above mass inventory in place, we finally construct the dynamical model of the stellar disc of FGC 2366 using the stellar dynamical code "AGAMA". To identify the key physical mechanisms responsible for the superthin vertical structure, we carry out a Principal Component Analysis of the data corresponding to all the relevant dynamical parameters and $a/b$ for a sample of superthin and extremely thin galaxies studied so far. We note that the first two principal components explain 80$\%$ of the variation in the data, and the significant contribution is from the compactness of the mass distribution, which is fundamentally responsible for the existence of superthin stellar discs.

\end{abstract}


\section{Introduction}
Superthin galaxies are bulgeless edge-on disc galaxies with low surface brightness, i.e., with central surface brightness in the $B$-band $\rm \mu_{B}(0)>22.7\, mag\,arcsec^{-2}$ \citep{bothun1997low, mcgaugh1996number} and characterized 
by large values of the major-to-minor axes ratio $a/b$ i.e., $10<a/b<22$ \citep{1999BSAO...47....5K}. For a review of properties of superthin galaxies, see \cite{2000bgfp.conf..107M} and \cite{kautsch2009edge}. Superthin galaxies have been studied in various observational surveys like the SDSS 
\citep{kautsch2006catalog,bizyaev2017very, 2021ApJ...914..104B}. The Revised Flat Galaxy Catalogue (RFGC), sourced from the Palomar Observatory Sky Survey (POSS-II) survey, is the primary catalogue of edge-on disc galaxies \citep{ 1999BSAO...47....5K}.
It contains 4,444 edge-on galaxies with $a/b > 7$, 1150 galaxies with an $a/b$ $>$ 10, and only 6 extremely thin galaxies with  $a/b$ $>$ 20, indicating the paucity of extremely thin galaxies. In Figure 1, we show the distribution of the $a/b$ ratio of the RFGC population. We see that the extremely thin galaxies constitute the tail of the distribution of the RFGC. The term \emph{superthin} was first used by \cite{goad1981spectroscopic} for galaxies with $a/b$ in the range 9 - 20. \cite{karachentseva2016ultra} classify
the galaxies in the RFGC with $a/b >10$ as \emph{ultra-flat} galaxies and use the term interchangeably with the \emph{superthin} galaxy, which is also the nomenclature adopted in the review by \cite{kautsch2009edge}. \cite{matthews1999extraordinary} use the ratio of the vertical exponential scaleheight
 of the stellar disc-to-the radial disc scalelength $h_{z}/R_{d} \leq 0.1$ as the definition of \emph{superthin} galaxies. In this work, we use the term \emph{extremely thin galaxies} to refer to RFGC galaxies with $a/b > 20$ and the term \emph{superthin} for $10 < a/b <20 $.  Although superthin galaxies are ubiquitous, as indicated in Figure 1, curiously, the number of superthin and extremely thin galaxies produced in the $\Lambda$ cold dark matter $(\Lambda CDM)$ simulations is not significant. In a recent study, \cite{haslbauer2022high}, measured the sky-projected aspect ratio distribution in the $\Lambda CDM$ simulations:  IllustrisTNG \citep{pillepich2018simulating}, and EAGLE \citep{2015MNRAS.446..521S} and found that these simulations are deficient in galaxies with intrinsically thin discs. (Also see \cite{vogelsberger2014introducing} and \cite{2017MNRAS.467.2879B}). 

\begin{figure} 
\resizebox{70mm}{50mm}{\includegraphics{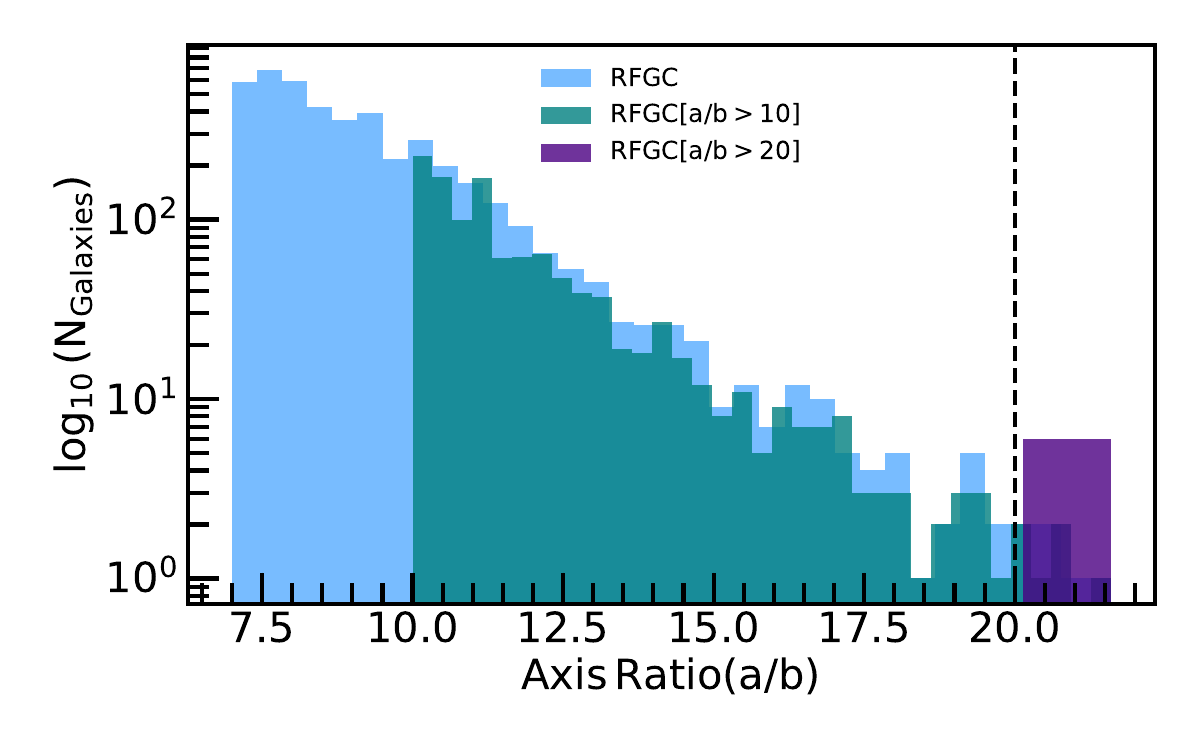}} 
\caption{Distribution of the RFGC galaxies with respect to the major-to-minor axes ratio $a/b$ of their stellar discs. The vertical line demarcates the superthin galaxies (Left) from extremely thin galaxies (Right).}   
\end{figure}

Our current understanding of the superthin galaxies is mostly derived from the detailed dynamical modelling of a limited sample of superthin galaxies using stellar photometry available in the literature and \HI{} 21cm radio-synthesis observations. Most of these galaxies have $ 10 < a/b < 16$, i.e UGC 7321  \citep{o2010dark,banerjee2013some}, IC5249 \citep{van2001kinematics,o2010dark}, IC 2233 \citep{matthews2008corrugations}, 
FGC 1540 \citep{kurapati2018mass}. \HI{} 21cm observations in edge-on galaxies offer a two-fold constraint on the net gravitational potential of the galaxy, i.e the rotation curve constraining the radial derivative, and the vertical gas thickness the vertical derivative of the same \citep{1995AJ....110..591O,narayan2002origin}. Using disc dynamical stability arguments, \cite{zasov1991thickness} showed that a massive dark matter halo stabilizes a superthin disc. Besides, \cite{ghosh2014suppression} showed that the prototypical superthin galaxy UGC 7321 is stable against local, axisymmetric and non-axisymmetric perturbations due to its dark matter halo. \cite{ banerjee2010dark} argued that UGC 7321 is dominated by dark matter at all galactocentric radii, and \cite{banerjee2017mass}  found that superthin galaxies have 
compact dark matter halos, i.e., $R_{c}/R_{d} < 2$ where $R_{c} $ is the core radius of a pseudo-isothermal dark matter halo and $R_{d}$ the exponential stellar disc scale length. Besides, \cite{2013MNRAS.431..582B} argued that a compact dark matter halo plays a decisive role in regulating the thin vertical structure of the stellar disc. Further, the razor-thin vertical structure of the superthin galaxies indicates the presence of an ultra-cold stellar disc. This may also be the reflection of a highly anisotropic stellar velocity ellipsoid with relatively 
less heating in the vertical as compared to the radial direction \citep{quinn1993heating, khoperskov2003minimum, purcell2010heated,gentile2015disk, grand2016vertical}. However, the vertical component of the stellar velocity dispersion of superthin galaxies is not directly measurable from the observations, given their edge-on orientation. Using a sample of five superthin galaxies with  $a/b < 16$, \cite{10.1093/mnras/stab155} theoretically determined the stellar vertical velocity dispersion as a function of the radius using the observed
scaleheight of the stellar disc as one of the constraints, and found the ratio of the vertical velocity dispersion to the radial velocity dispersion lying between 0.3 – 0.5 in comparison to a value of $\sim$ 0.5 as seen in ordinary disc galaxies. Also, the ratio of the vertical velocity dispersion to rotational velocity was found to be comparable to the Galactic thin disc. Finally, basic physics suggests that a high value of the planar-to-vertical axes ratio of the stellar disc could be the outcome of a high value of the specific angular momentum of the disc. Interestingly, \cite{jadhav2019specific}  found that superthin galaxies have a relatively high specific angular momentum at a given mass compared to bulgeless ordinary disc galaxies as given in \cite{romanowsky2012angular}, indicating the possibility of a high value of specific angular momentum driving the superthin stellar discs. 

In this paper, we choose the thinnest galaxy, FGC 2366, for its first-ever \HI{} 21cm radio-synthesis observations using the Giant Meterwave Radio Telescope (GMRT), followed by modelling the structure and kinematics of \HI{} using the 3D tilted ring modelling software Fully Automated TiRiFiC (FAT, \cite{kamphuis2015automated}). We model the \HI{} rotation curve and \HI{} surface density, which, along with stellar photometry available in the literature, was used for mass modelling and detailed dynamical modelling of these galaxies. We focus on four dynamical quantities that may be instrumental in driving the strikingly high planar-to-vertical axes ratios in these galaxies, namely, the 1) Dark matter density profile, 2) Stellar vertical velocity dispersion, and the ratio of the vertical-to-radial stellar velocity dispersion, 3) Disc dynamical stability against local axisymmetric perturbations 4) Specific angular momentum of the disc. We then compare and contrast the above physical parameters of FGC 2366 with those of the previously studied extremely thin galaxy FGC 1440, other superthins, and ordinary disc galaxies. Finally, we do a Principal Component Analysis of a set of dynamical parameters corresponding to our sample of superthins and extremely thin galaxies to identify the key dynamical mechanism responsible for the superthin vertical structure of their stellar discs. The rest of the paper is organised as follows; in \S 2, we introduce the target FGC 2366; in \S 3, we describe the \HI{} observations, data reduction, analysis, and 3-D tilted ring modelling, and in \S 4, the optical photometry followed by the dynamical modelling and conclusions in \S 5 and \S 6 respectively.

\section{FGC 2366}
FGC 2366 is an extremely thin galaxy from the RFGC \citep{ 1999BSAO...47....5K}. 
The ratio of the major-to-minor axes for FGC 2366 is equal to 21.6, which makes it the thinnest or the flattest known galaxy. 
FGC 2366 was observed as part of the survey of 472 late-type, edge-on spiral galaxies using Nancy Radio Telescope and the Green Bank Telescope by 
\cite{matthews2000h}; an integrated \HI{} flux density equal to  5.93 Jy \kms{}, and velocity width at 50$\%$ of peak flux $(W_{50})$ equal to 188 
\kms were reported. The basic properties of FGC 2366 are presented in Table 1.

\begin{table}
\begin{minipage}{110mm}
\hfill{}
\caption{Basic astrophysical parameters of FGC 2366}
\begin{tabular}{|l|c|}
\hline
\hline
Parameter& Value \\
\hline    

RA(J2000)$^{\color{red}(1)}$\footnote{Right ascension}   &  $22^h 08^m 03\fs62$ \\
Dec(J2000)$^{\color{red}(1)}$\footnote{Declination}  & $-10\degr19\arcmin59\farcs1$ \\
Hubble type$^{\color{red}(1)}$ & Sd \\
$i^{\color{red}(3)}$ \footnote{Inclination}   & $90^{\circ}$\\
Distance$^{\color{red}(4)}$ &32.95 Mpc\\
a/b$^{\color{red}(1)}$\footnote{Major axis to minor axis ratio  } & 21.6  \\
log (M$_{HI}$)$^{\color{red}(2)}$\footnote{\HI{} mass} & 9.38\\
W$_{50}$$^{\color{red}(5)}$\footnote{Spectral line width at 50$\%$ of the peak flux density}  & 204.5\\

\hline
\end{tabular}
\hfill{}
\label{table: table 1}
\end{minipage}
\begin{tablenotes}
\item \textcolor{red}{(1)}: \cite{1999BSAO...47....5K} 
\item \textcolor{red}{(2)}: \cite{matthews2000h} 
\item \textcolor{red}{(3)}: \cite{2014A&A...570A..13M} 
\item \textcolor{red}{(4)}: \cite{kourkchi2020cosmicflows}
\item \textcolor{red}{(5)}: \textcolor{blue}{This Work} 
\end{tablenotes}
\end{table}

\section{\HI{} 21cm radio-synthesis observations}
We observed FGC 2366 using GMRT on 25 August 2019, for 9 hours, including overheads and with 27 working antennae. 
The source was observed for 7 hours through 14 scans of half an hour each, interspersed by 14 scans of the phase calibrator 2136+006 for 5 minutes each. 
The flux calibrator 3C286 was observed at the beginning, and another, 3C48, at the end of the observation for half an hour. We have observed the central frequency of 1406.9 MHz in GSB mode with 512 channels with a total bandwidth of 4.2 MHz and a spectral resolution of 8.14 kHz. A velocity resolution of 1.7 \kms{} allows us to model the dispersion of the gas as well as to cover the full velocity extent of the galaxy. \HI{} emissions in multiple channels allows us to construct realistic 3D model of the galaxy using FAT, see \S 3.2. We use the Common Astronomy Software Application (CASA) \citep{mcmullin2007casa} for data reduction. We start by flagging off the bad antennae (E04, E05, E06) from our data set 
and then visually inspect and flag the data corrupted by radio frequency interference (RFI). We then follow the usual procedure for 
cross-calibration before splitting off the target from the measurement set. Thereafter, we make a dirty cube to locate the channels with spectral line emission and flag off those channels to constitute a continuum-only measurement set. Next, we perform 5 rounds of phase-only self-calibration and four rounds of amplitude-phase self-calibration, each time applying the self-calibration solutions to the measurement set. The signal-to-noise ratio (S/N) saturates after about 5 rounds of phase-only self-calibration. During amplitude-phase self-calibration, the image quality initially deteriorates but improves after the second round, and the S/N ratio saturates after the third round. We apply the final amplitude-phase self-calibration table to the target-only measurement set. We then subtract the continuum, excluding the channels containing the spectral line, with a zeroth order interpolation excluding the spectral channels. We then make a dirty cube using the continuum subtracted measurement set and create a mask using the Source Finding Algorithm (SoFiA: \cite{serra2015sofia}). Finally, we cleaned the dirty cube using a mask made using SoFiA to the final data cube. Upon experimenting, we found that $'briggs'$ weighing scheme in the CASA task 'tclean' with robustness parameter equal to 0.5 and a uvtaper equal to 12k$\rm \lambda$ gives the best compromise between sensitivity and resolution. The details of the observations and the deconvolved image characteristics are summarized in Table 2.

\begin{table}
\caption{ \HI{} observations of FGC 2366: Technical specifications}
\begin{tabular}{|l|c|}	
\hline
\hline
(a) Observing Setup&\\
\hline
Parameter& Value \\
\hline    
\hline
Observing Date                     & 25 August 2019\\
Phase center,$\alpha$(J2000)       & $22^h 08^m 03\fs62$\\
Phase center,$\delta$(J2000)       & $-10\degr19\arcmin59\farcs1$\\
Total on-source observation time   &  $5.5$ hrs\\
Flux  calibrator                   & 3C286, 3C48  \\
Phase calibrator                   & 2136+006\\
Channel Width                      & 8.14 kHz\\
Velocity separation                & 1.7 \kms\\
Central frequency                  & 1406.9 MHz\\
Total bandwidth                    & 4.2 MHz   \\
\hline
(b) Deconvolved Image Characteristics&\\
\hline
Weighing                           & robust\\
Robustness parameter               & 0.5\\
Synthesized beam FWHM              & $13.4\farcs \times 11.7\farcs$\\
Synthesized beam position angle    & $39.4^{\circ}$\\
rms noise in channel               & 0.39 mJy/beam\\
\hline
\end{tabular}
\hfill{}
\label{table: table 2}
\end{table}

\subsection{Data Analysis}
In Figure 2, we present the global \HI{} profile of FGC 2366. We fit the observed profile using the busy-function \citep{westmeier2014busy} and find
that the integrated flux is equal to 5.4 Jy \kms and the peak \HI{} flux equal to 34 mJy. The integrated flux obtained by fitting the busy function to the \HI{} spectrum is comparable to the flux measured directly from the data cube using the CASA task \textit{SPECFLUX}. The integrated flux gives us a \HI{} mass of galaxy equal to $\log_{10}(M_{HI}/M_{\odot})=9.1$. We find that the $(W_{20})$ and the $(W_{50})$ velocity widths are 214 \kms and 205 \kms, respectively. The results obtained by fitting the busy function are shown in Table 3. In Figure 3, we show an integrated column density (Moment 0) map and the velocity field (Moment 1) for FGC 2366. We have overlaid the column density map on the POSS-II  optical image of FGC 2366. We find that the Moment 0 map is smooth and unperturbed, and also, we do not find any enhanced emission from either the centre or the edge of the \HI{} disc. Also, we note that the \HI{} distribution and velocity field is similar to other superthin galaxies; for example, see \cite{kurapati2018mass}. 
In Figure 4, we present the images of channels containing \HI{} emission overlaid on the POSS-II optical image. For convenience, we have shown every fourth channel. We observe that the \HI{} emission close to the systemic velocity and away from the systemic velocity lie along a flat plane. Only in the edge channels does the \HI{} emission on both the approaching and the receding sides show a slight deviation away from the plane.

\begin{table}
\centering
\caption{Busy function fit to \HI{} Global profile}
\begin{tabular}{|c|c|c|c|c|}
\hline
\hline
$V^{\textcolor{red}{(a)}}_{0}$ & $W^{\textcolor{red}{(b)}}_{50}$ & $W^{\textcolor{red}{(c)}}_{20}$ & $F^{\textcolor{red}{(d)}}_{peak}$ & $F^{\textcolor{red}{(e)}}_{int}$ \\
\kms{} & \kms{} & \kms{} & mJy & $\rm Jy\, \kms{}$ \\
\hline
$2844 \pm 2.4$ & $204.5 \pm 2.9$ & $214.4 \pm 3.9$ & $33.9 \pm 0.3$ & $5.4 \pm 0.2$ \\
\hline
\end{tabular}
\label{table:table3}
\begin{tablenotes}
\item $\textcolor{red}{(a)}$: Frequency centroid of the \HI{} line.
\item $\textcolor{red}{(b)}$: Spectral line width at 50$\%$ of the peak flux density.
\item $\textcolor{red}{(c)}$: Spectral line width at 20$\%$ of the peak flux density.
\item $\textcolor{red}{(d)}$: Peak of the \HI{} flux density.
\item $\textcolor{red}{(e)}$: Integrated \HI{} flux.
\end{tablenotes}
\end{table}

\begin{figure} 
\resizebox{70mm}{50mm}{\includegraphics{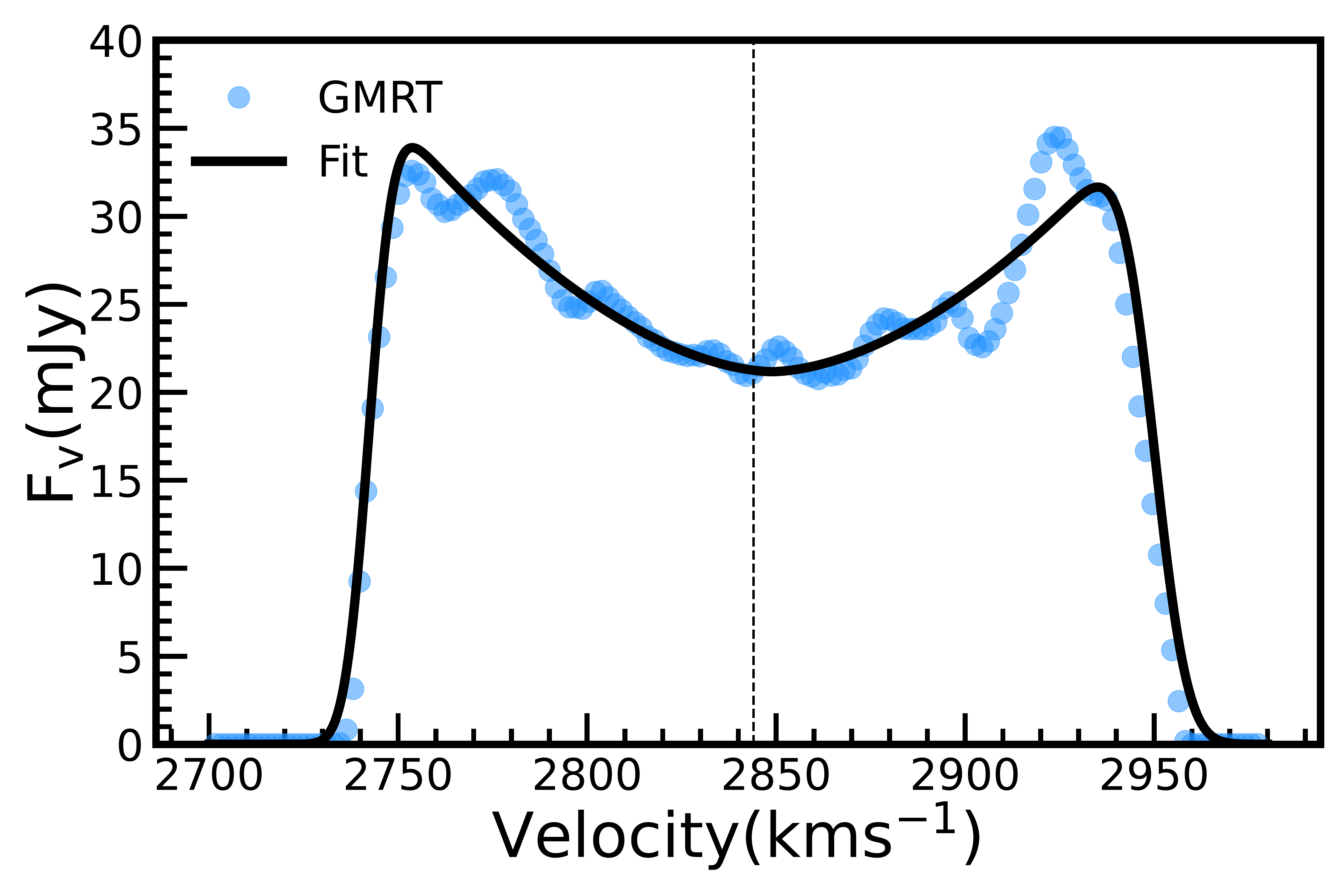}} 
\caption{ Global \HI{} profile of FGC 2366 derived from our GMRT radio-synthesis observation, with the best-fitting busy function, over-plotted.
The vertical dashed line indicates the systemic velocity.}
\end{figure}

\begin{figure*}
\resizebox{180mm}{60mm}{\includegraphics{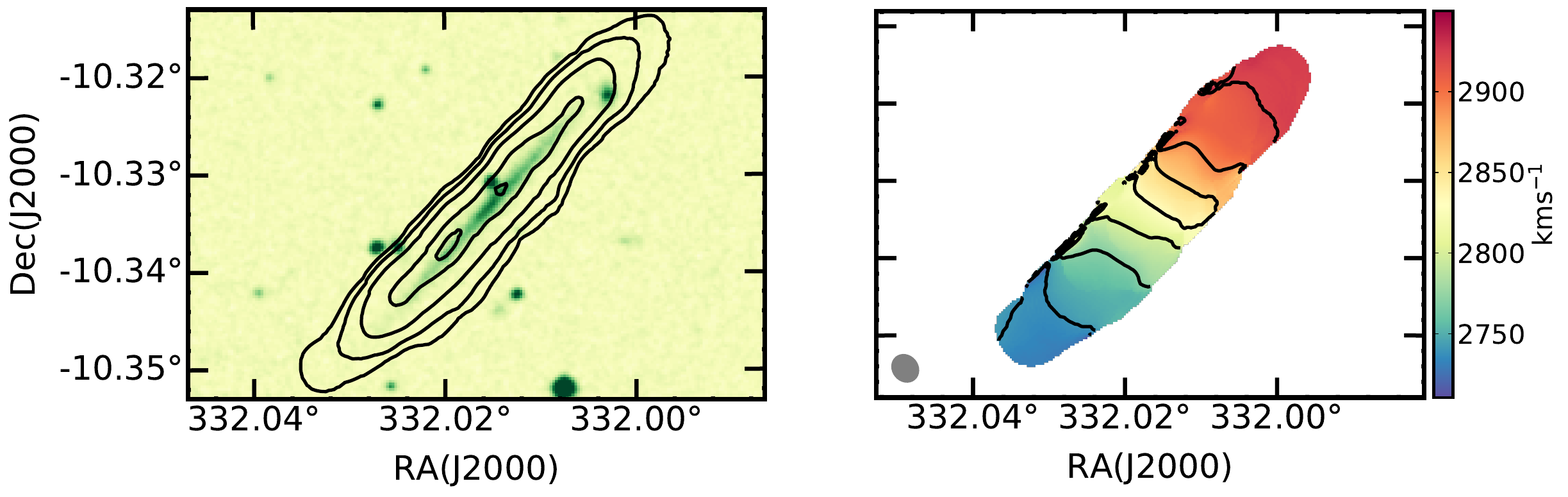}} 
\caption{Moment 0 [Left] and Moment 1 [Right] maps derived from the observed \HI{} data cube. The data contours are shown in black. 
The contours levels in the moment 0 maps at [ 4.0, 8.0, 16, 32, 40]$\times$ $10^{20}$ atoms cm$^{-2}$  and the contours in the moment 1 map start at 2951 \kms and end at 2710 \kms  
increasing by 26 \kms.}
\end{figure*}

\begin{figure*}
\resizebox{185mm}{175mm}{\includegraphics{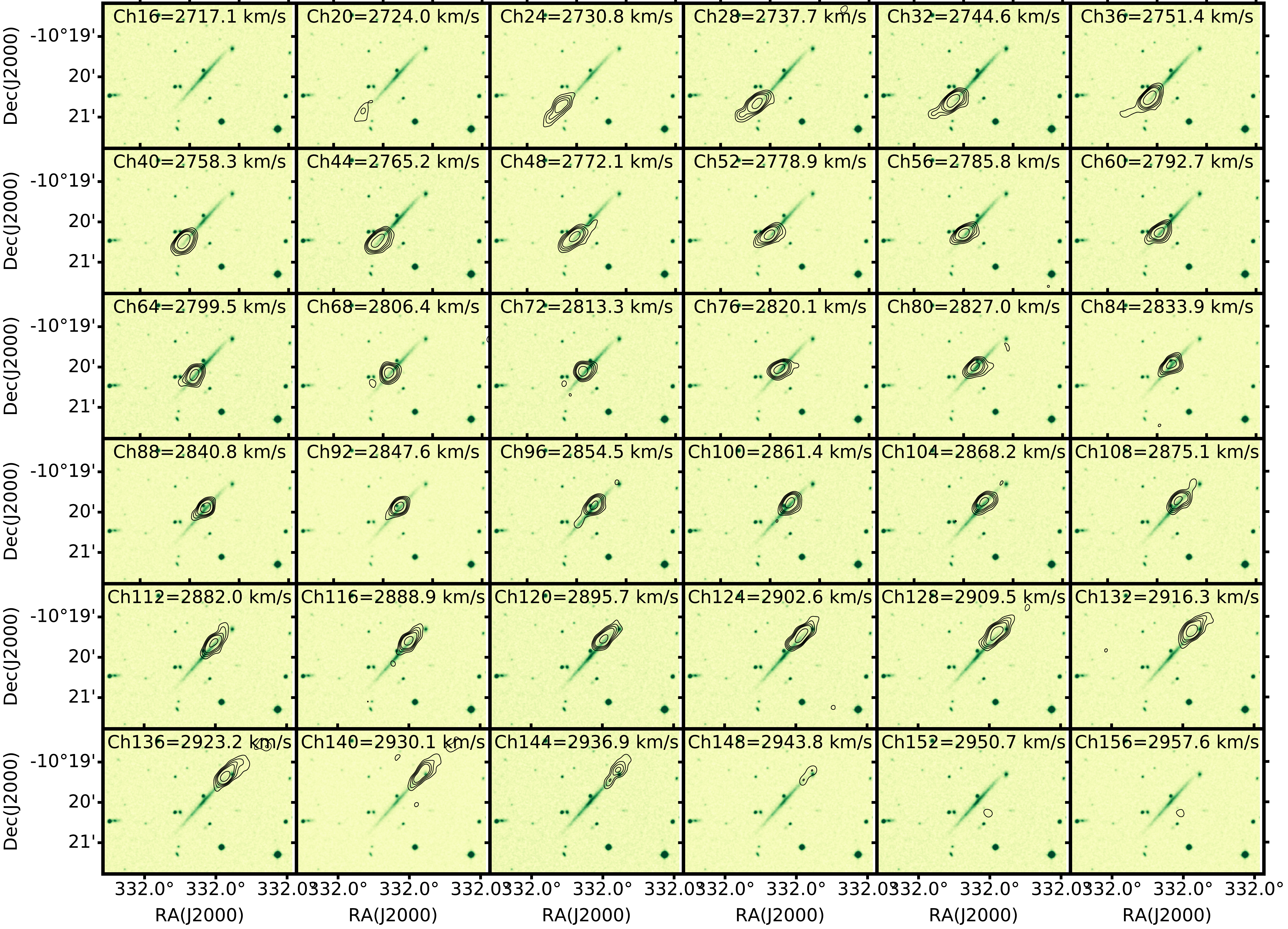}} 
\caption{Channel maps showing \HI{} emission from FGC 2366, each panel is separated by four channels. The \HI{} emissions are overlaid on the POSS-II optical image. 
The contour levels are at [3, 4, 5 ,6, 9]$\times$ 0.8 mJy beam$^{-1}$}	
\end{figure*}

\subsection{3D-Tilted Ring Modeling}
We use the publicly available automated 3D-tilted ring modelling software Fully Automated TiRiFiC (FAT) \citep{kamphuis2015automated}, which is a GDL wrapper around 
the tilted ring modelling software TiRiFiC \citep{2007A&A...468..731J}. TiRiFiC generates model data cubes from the tilted ring parametrization of the rotating disc and fits the model data cubes directly to the observed data cubes. Fitting the observed data cubes directly using TiRiFiC allows us to accurately derive the structural and kinematic properties of a completely edge-on galaxy like FGC 2366. This is otherwise not feasible using conventional 2D methods, which rely on the velocity field for the derivation of structural and kinematic properties and therefore suffer from effects arising from projection and beam smearing. The 2D velocity field is typically affected by beam smearing that affects the derivation of the rotation curve in the inner regions of the galaxy. In contrast, projection effects do not allow one to derive a unique velocity field for completely edge-on galaxies as the line-of-sight intersects the disc multiple times. These problems are circumvented by fitting directly to the observed data cube rather than fitting to derived data products like 2D-velocity fields or PV diagrams. FAT uses the \HI{} data cube as input and automates the fitting of the following free parameters: {1) surface brightness 
2)  position angle, 3) inclination, 4) rotational velocity, 5) scaleheight 6) intrinsic dispersion, and 7) central coordinates: right ascension, declination, and systemic velocity. Each of the free parameters can be fitted at every radial point independently.
We divide the \HI{} disc into two halves and fit 19 semi-rings across each half, with each semi-ring being 0.6 times the major axis beam size, and model the vertical gas distribution using the $\rm sech^{2}$ profile. The best-fitting values obtained by fitting 3D models using FAT are shown in Table 4; the rotation curve, velocity dispersion, and the surface brightness profiles thus obtained are presented in Figure 5. The red and black lines indicate the surface brightness profiles of the approaching and the receding sides, respectively. From the \HI{} surface brightness profile, we note 
the presence of the central \HI{} hole, which could be due to supernova-driven outflows or conversion of the dense atomic gas in the centre into molecular hydrogen gas. In Figure 5, we note that the surface brightness peaks have a similar value but occur at a different radius. From the major axis PV diagram, we note that the approaching side has a slightly higher surface brightness than the receding. Thus, despite symmetric surface brightness maps, 
the asymmetry in the velocity space is reflected in radially integrated surface brightness profiles. Also, we find the 3D model of FGC 2366 constructed using FAT accurately describes the \HI{} distribution except for the asymmetry at the highest contour, see Figure 7. 

In Figure 6, we overlay the rotation curve derived using the 3D model on the major axis PV diagrams. We observe that the shape of the rotation curve derived from the 3D modelling agrees with the shape of the mid-plane emission in the PV diagram along the major axis. Further, from the PV diagram along the major axis, we note that the galaxy has a similar rotation curve for the approaching and the receding side. We observe that the central velocity estimated by FAT differs from that obtained by fitting the busy function as the FAT fits the 3D data cube directly. However, we may note here that we fit the busy function to the spectrum derived from the 3D cube, which is a 1D representation of the data. Further, the \HI{} spectrum, as given in Figure 2, shows slight asymmetry, which could be a possible reason for the difference in the value of the central velocity obtained by FAT and busy function fitting. Further, we over-plot the emissions observed from channels at 2844 \kms{} (estimated from the 1D spectrum) and 2832.5 kms$^{-1}$ (estimated from the 3D data cube using FAT). We find that the emission observed at 2832.5 kms$^{-1}$ coincides with the centre of the optical image. (See Figure 20 in Appendix A.) In Figure 6, we plot the systemic velocity estimated by FAT on the major axis PV diagram. We can clearly see that the systemic velocity estimated by FAT coincides with the centre of the galaxy. Throughout this work, we will use 2832.5 \kms{} as the systemic velocity. Further, from Figure 6, we can see that the approaching side is brighter than the receding  side, possibly indicating kinematic asymmetry, which is driving the difference between the values of systemic velocity estimated from 3D data cube using FAT and from the 1D spectrum.

We define the uncertainties on the observed rotation curve following \cite{swaters2009rotation} and \cite{de2008high}. We quantify the two main sources of uncertainties on the rotation curve; one arising due to the difference in the rotation velocities between the approaching and the receding sides and the second due to the spectral resolution of our observation. We define the final uncertainty by adding in quadrature the spectral resolution and one-fourth of the difference of the rotation velocity on the approaching and the receding side. Since we find a similar rotation velocity on the approaching and receding sides, the errors on our rotation curve are primarily the spectral resolution of our observation.

\textbf{Quality of Fits:}
In Figure 7, we compare the Moment 0 and Moment 1 map derived from the input and model data cube. The black contours depict the data, and the crimson contours show the 3D model of the \HI{} distribution generated by TiRiFiC. In Figure 7, we plot the velocity contours separated by 26 \kms{} so 
that we can have sufficient contours at the centre and away from the centre to compare the model and the data amenably. The model contours closely follow the observed \HI{} distribution. The moment maps contain information about the surface brightness distribution and velocity separately, whereas the observed data cube has a three-dimensional structure. Therefore it is necessary to compare the minor axis PV diagrams, which preserve the 3D structure of the observed data cube. In Figure 8 [top panel], we compare the minor axis PV diagrams derived from the input data cube and the model data cube, and in the bottom panel, we present the residual plots. We have spatially smoothed the data cube for making the minor axis PV diagrams to allow for overall comparison between model and data amenable.
Our model data cube reproduces the overall observed structure of the \HI{} emission, and the difference between the model and the data arises only at velocities away from the galaxy close to 3$\sigma$.

\begin{table}
\begin{minipage}{110mm}
\hfill{}
\caption{Best-fitting model derived by FAT }
\begin{tabular}{|l|c|}
\hline
Parameter&    Values  \\
\hline    
$\rm X_{o}$    \footnote{Right ascension}            &   $22^h 08^m 03\fs62$               \\
$\rm Y_{o}$    \footnote{Declination}            &    $-10\degr19\arcmin59\farcs1$                 \\
$\rm i$        \footnote{Inclination}            &    $87.1 ^{\circ}\, \pm \, 4^{\circ}$            \\
$\rm V_{sys}$  \footnote{Systemic velocity}         &    2832.5 \kms               \\
PA             \footnote{Position angle}           &    $317.1^{\circ} \, \pm \, 2^{\circ} $         \\
$\rm h_{z}$      \footnote{Scaleheight of the \HI{} disc}         &    6.7 \farcs            \\
\hline
\end{tabular}
\hfill{}
\label{table: table 4}
\end{minipage}
\end{table}

\begin{figure*}
\resizebox{185mm}{45mm}{\includegraphics{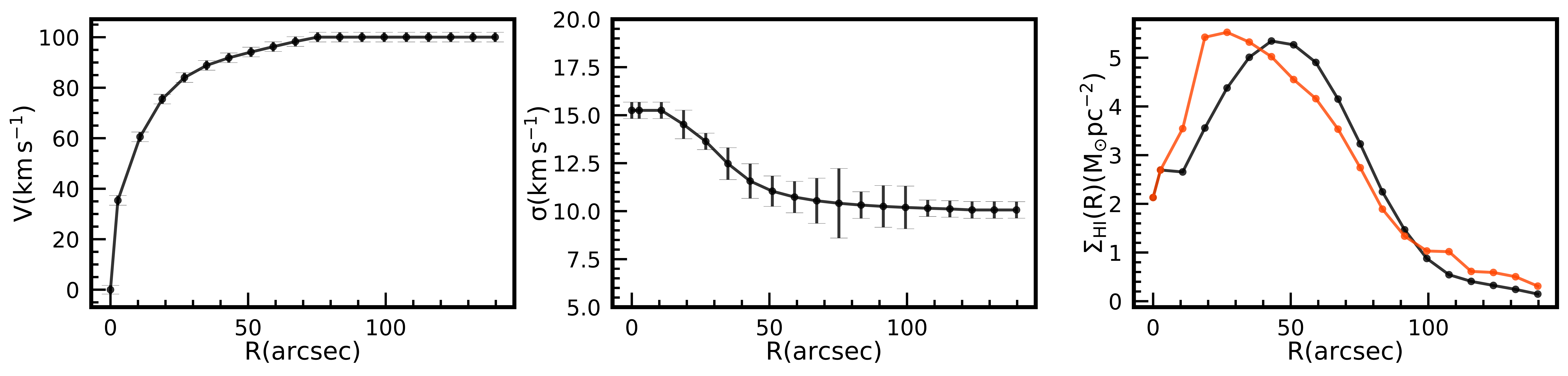}} 
\caption{Best-fitting model derived using the 3-D tilted modelling of the observed \HI{} data cube. [Left Panel] Rotational velocity [Middle Panel] \HI{} dispersion [Right Panel] \HI{} surface density as a function of galactocentric radii. The red and black lines in the right panel indicate the 
surface brightness profile of the approaching and receding sides, respectively.}	
\end{figure*}

\begin{figure}
\resizebox{90mm}{46mm}{\includegraphics{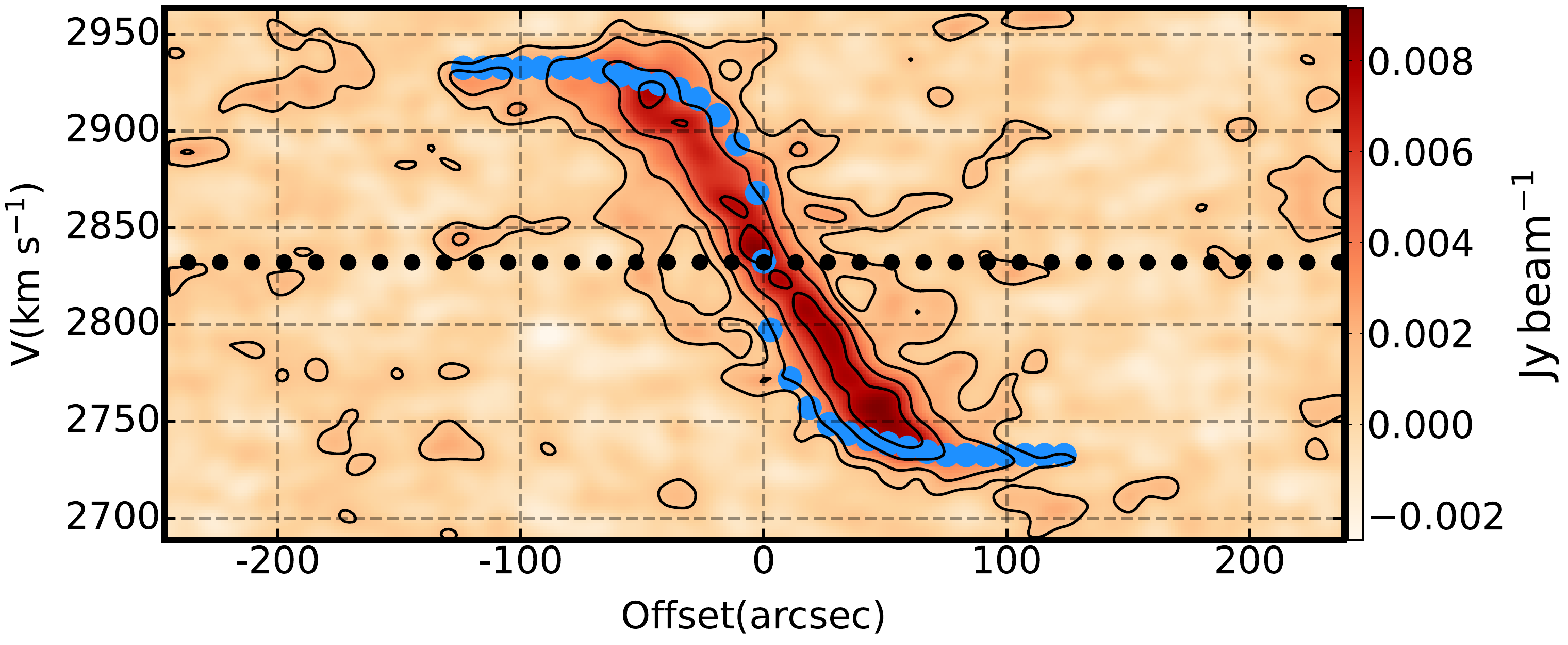}} 
\caption{PV diagrams of FGC 2366 derived along the major axis. 
We have overplotted the rotation curve derived from the 3D tilted ring modelling [blue points]. The contour levels are at  [1.5,3, 6, 9, 12]$\times$ 0.8 mJy beam$^{-1}$. The black dotted line indicates the systemic velocity estimated by fitting 3D models to the observed data cube.}	
\end{figure}

\begin{figure*}
\begin{center}
\resizebox{180mm}{60mm}{\includegraphics{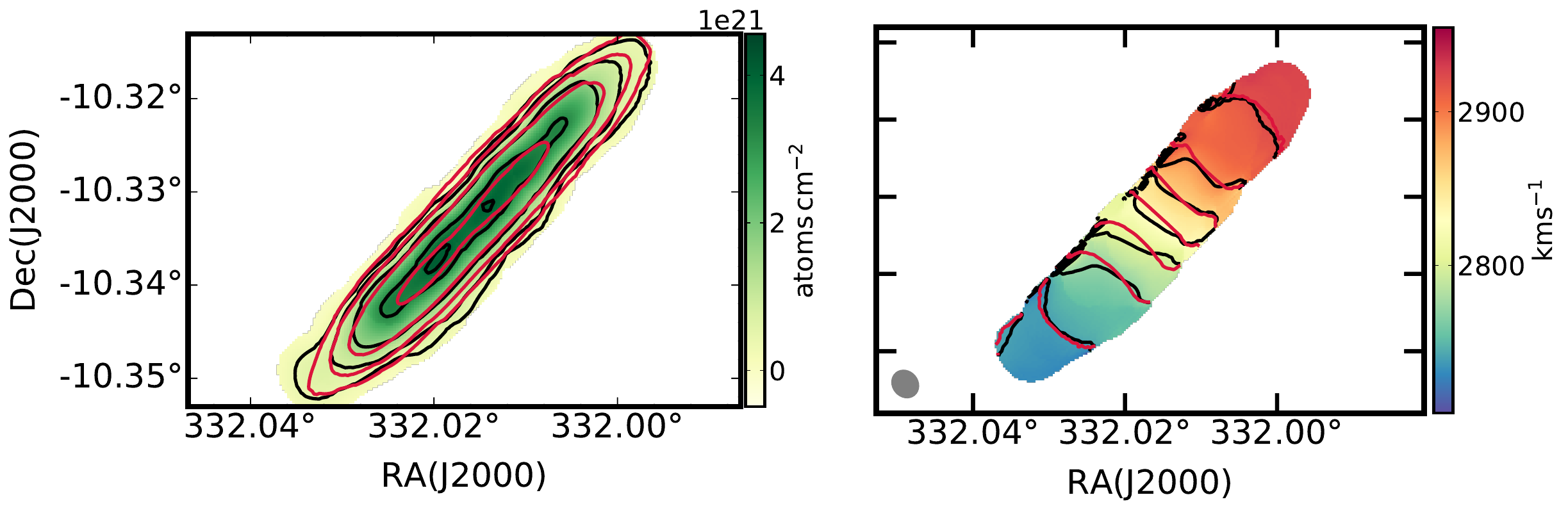}} 

\end{center}
\caption{Moment 0 and Moment 1 map derived from the observed and the model \HI{} data cube. The model contours
are shown in crimson, and the data contours are shown in black. The contours levels in the moment 0 maps are at [ 4, 8, 16, 32, 40]$\times$ $10^{20}$ atoms cm$^{-2}$  
and the contour in the Moment 1 map starts at 2951 \kms and ends at 2710 \kms, increasing by 26 \kms.}	
\end{figure*}

\begin{figure*}
\begin{center}
\begin{tabular}{cc}
\resizebox{180mm}{40mm}{\includegraphics{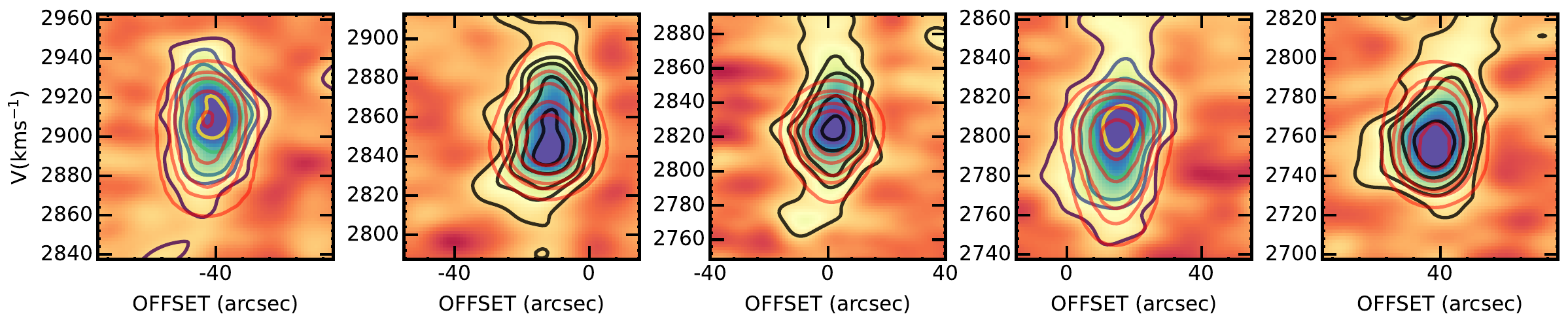}} \\
\resizebox{180mm}{40mm}{\includegraphics{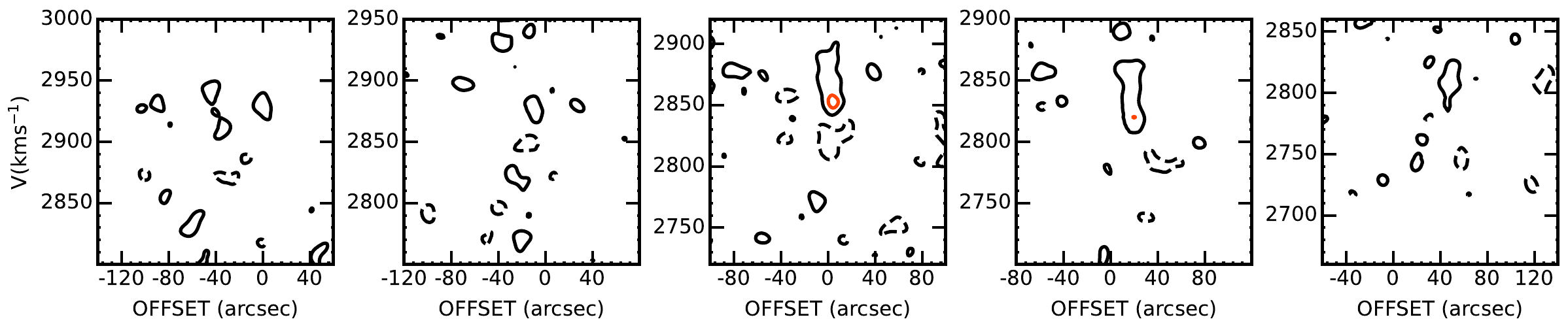}} 
\end{tabular}
\end{center}
\caption{In the top panels, we show the PV diagrams of the observed data cube and the model data cube along the minor axis. The contour levels are 
at  [1.5, 3, 4, 6, 8]$\times$ 1.0 mJy beam$^{-1}$. The black contours in the top panel indicate the data, and the red coloured 
contours correspond to the model. In the bottom panels we plot the residuals corresponding to the minor axis PV diagrams were obtained by subtracting the model from the data.}	
\end{figure*}

\section{Optical photometry}
We derive the structural parameters for FGC\,2366 in the $g$, $r$, and $z$ bands using optical images from the DESI Legacy Imaging Surveys DR9 \citep{2019AJ....157..168D}. 
This galaxy does not exhibit an apparent dust lane or any traces of dust, even in the bluest optical bands. Therefore, to a first approximation, we can neglect the internal extinction in this galaxy and roughly estimate its structural parameters for all three available bands. We use the publicly available photometric decomposition package IMFIT \citep{erwin2015imfit} and employ the ExponentialDisk3D function with the galaxy inclination parameter. We fit the light distribution keeping the central surface brightness, radial scalelength, vertical exponential scaleheight, and inclination as free parameters. 
The results of our decomposition for all three bands are shown in Table 5. We estimate the total magnitudes of the galaxy model 
in the $g$, $r$, and $z$ bands, corrected for the Galactic extinction from \cite{schlafly2011measuring}, to be 15.44, 15.01, and 14.76, respectively. From Table 5, we can see that the radial scalelength decreases with wavelength, which conforms with the inside-out formation scenario. Unlike the disc scalelength, we find that the disc scaleheight is slightly smaller in the redder 
z-band than the g-band and the r-band, but the variation in the scaleheight is comparable within the error bars.
The scale height in all three bands does not change within the error bars. The model and data profile along the major axis in the z-band are shown in Figure 9.\\

\noindent \textbf{Mass-to-light ratio:}\\
We derive the mass-to-light ratio in the z-band in order to convert the surface brightness profile in $L_{\odot}/pc^{2}$ unit to the surface density profile in $M_{\odot}/pc^{2}$ using the calibration given in \cite{bell2003optical}. The empirical calibration between the colour and the mass to the light ratio in a given band is given as  $\log_{10}(M/L)=a_{\lambda} +b_{\lambda} (Color)$. We use the g-z magnitudes to derive the colour and the values of $a_{z}$ and $b_{z}$ tabulated in \cite{2003ApJS..149..289B} and compute the mass-to-light ratio in the $z$ band. The details are presented in Table 6. \\ 
\noindent We use the $z$-band photometry for constructing the mass models as it is a better tracer of the total stellar luminosity than the $g$ and the $r$ bands. Using the values of the $a_{\lambda}$ and $b_{\lambda}$ corresponding to the $z$-band photometry, we find an $M/L$ value equal to 1.1 using a scaled Salpeter initial mass function (IMF). \cite{palunas2000maximum}, using I-band photometry for a sample of 75 spiral galaxies, find M/L= $2.4\pm 0.9$. Using K-band photometry \cite{mcgaugh2014color} show that the M/L remains fairly constant at 0.6 for spiral galaxies.

\begin{figure}
\resizebox{60mm}{50mm}{\includegraphics{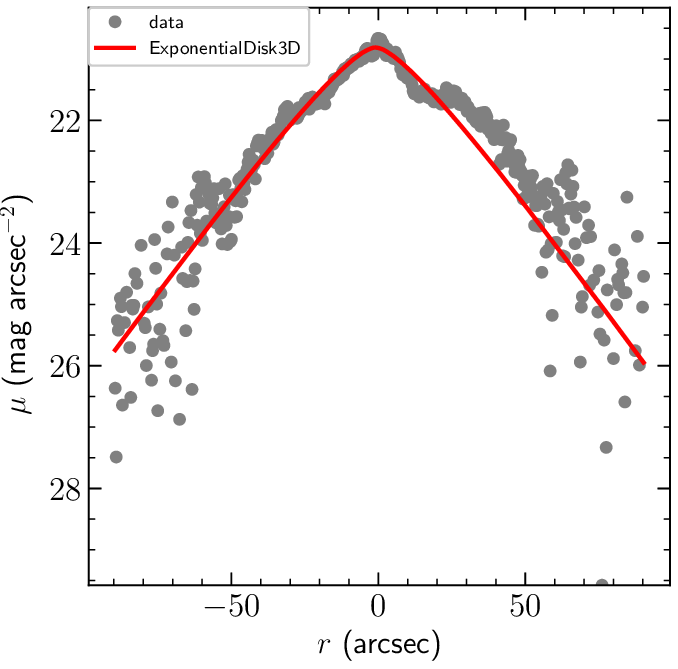}} 
\caption{The model and data profiles along the major axis of FGC 2366 for the $z$-band optical image. The light distribution is modelled using the ExponentialDisk3D function, keeping the surface brightness, scaleheight, scalelength, and inclination as the free parameters.}	
\end{figure}

\begin{table*}
\hspace{-6cm}
\begin{minipage}{110mm}
\hfill{}
\caption{Structural parameters derived from optical photometry of FGC 2366.}
\centering
\begin{tabular}{|l|c|c|c|c|}
\hline
\hline
Parameter                                 &               & 	               &               		&					 Description \\
                                          &g-band 	  &  r-band            &  z-band       		&             				\\
\hline
Total magnitude                           & 15.44         &   15.01            & 14.76          	&   					\\
$\mu^{edge-on}_{o}$                       & $21.58 \pm 0.02$&  $ 20.97\pm 0.02$    & $20.46 \pm 0.02$	        &	Edge-on surface brightness ($\rm{mag/arcsec}^{2}$)\\
$R_{d}$                                   & $20.4  \pm 0.2$ &   $17.9 \pm 0.1$     & $16.1  \pm 0.1$	        &	Disc scalelength ($\rm{arcsec}$)\\
$h_{z}$                                   & $2.0   \pm 0.1$ &   $2.0 \pm 0.1$      & $1.8   \pm 0.1$	       	&       Disc scaleheight ($\rm{arcsec}$)\\
$i$                                       & $89.3  \pm 0.1$ &   $89.0\pm 0.1$      & $88.6  \pm 0.1$        	&	(\rm{degrees}).\\
\hline 
\end{tabular}
\hfill{}
\label{table: table 4}
\end{minipage}
\end{table*}

\section{Dynamical Modelling}
\subsection{Mass Modeling}
In this section, we determine the contribution of the dark matter halo to the net gravitational potential by decomposing the total rotation curve into baryonic $(stars + gas)$ and dark matter components, respectively. The total rotation curve of the galaxy is obtained by adding in quadrature the circular velocity we would have seen if the net gravitational potential was due to each component alone.

\begin{table*}
\hspace{-6cm}
\begin{minipage}{110mm}
\hfill{}
\caption{Input parameters for the stellar rotation curve.}
\centering
\begin{tabular}{|l|c|c|}
\hline
\hline
Parameter                                 &Value         &Description \\
                                          &z-band                     \\
\hline
$\mu^{face-on(*)}_{o}$   &22.8&         Face-on surface brightness ($\rm{mag/arcsec^{2}}$)\\
$\Sigma_{o}$                              &21.5&         Surface density ($L_{\odot}/pc^2$)\\
$R_{d}$                                   &2.6&          Disc scalelength (kpc)\\
$h_{z}$                                   &0.29&         Disc scaleheight (kpc)\\
\hline 
Input parameters for mass-to-light ratio $M/L$ \\
\hline
$g-z$         &  0.7      &                         \\
$a_{\lambda}$ &  $-0.17$  &    \cite{bell2003optical}\\
$b_{\lambda}$ &  0.32     &     \cite{bell2003optical}\\
$M/L$  &  1.12     &     M/L ratio derived using scaled Salpeter IMF\\
$M/L$  &  0.8      &     M/L ratio derived using Kroupa IMF\\
\hline
\end{tabular}
\hfill{}
\label{table: table 5}
\end{minipage}
\begin{tablenotes}
\item  (*): The edge-on surface brightness has been converted to face-on surface brightness using 
$ \mu^{face-on}= \mu^{edge-on} + 2.5log( \frac{ R_{d} } {h_{z}})$ \citep{kregel2005structure}.
\end{tablenotes}
\end{table*}

\begin{equation}
V^{2}_{Total}= (M/L) V^{2}_{*} + V^{2}_{Gas} + V^{2}_{DM}
\end{equation}
where $M/L$ is the mass-to-light ratio , $V_{*}$, $V_{Gas}$ and $V_{DM}$ are the circular velocity due to 
the stars, gas, and dark matter potentials alone, respectively. The observed rotational velocity is determined using tilted ring modelling, as discussed in \S 3.2. We derive the circular velocity of the gas disc $V_{Gas}$ by modelling the gas disc as thin concentric rings using the GIPSY  task ROTMOD \citep{van1992groningen}. We use the average \HI{} surface density of the approaching and the receding side as a function of radius obtained from the tilted ring model (Figure 5, Third Panel) as the input parameter in ROTMOD. 
The gas surface density is scaled by 1.4 for helium and other metals. Similarly, we derive the circular velocity of the stellar disc using the structural parameters derived 
from z-band photometry; the structural parameters are given in Table 6. 
We model the dark matter distribution using the observationally motivated pseudo isothermal (PIS) dark matter halo  \citep{begeman1991extended, 1998A&A...336..878F} characterized by a constant density core and, also using the 
Navarro-Frenk-White (NFW) dark matter halo profile derived from the
cold dark matter (CDM) simulations \cite{navarro1997universal}. The rotation curve of the cored PIS halo is given by

\begin{equation}
 V(R)=\sqrt{4\pi G \rho_{0} R^{2}_{c} \bigg( 1- \frac{R_{c}}{R} \rm{arctan}(\frac{R}{R_{c}})\bigg)} 
\end{equation}
where $\rho_{0}$ is the central density of the halo and $R_{c}$ is the core radius. The rotation curve due to the cuspy NFW density distribution is,
\begin{equation}
 V(R)=V_{200} \sqrt{\frac{\rm{ln}(1+cx) -cx/(1+cx)}{x[\rm{ln}(1+c) -c/(1+c)]}}
\end{equation}
where $x=R/R_{200}$, $R_{200}$ is the radius at which the mean density of the dark matter halo is 200 times the critical density. $V_{200}$ is the rotation velocity  at $R_{200}$. 
The concentration parameter is defined as $c=R_{200}/R_{s}$, and $R_{s}$ is the characteristic scale radius of the NFW density profile. 

Finally, the likelihood
function is defined as $\rm exp(-\frac{\chi^{2}}{2})$, where $\chi^{2}$ is given by,
\begin{equation}
 \chi^{2} =\sum _{R} \frac{\bigg(V_{\rm{obs}}(R) - V_{\rm{Total}}(R) \bigg)^{2} }{V^{2}_{\rm{err}}(R)}
\end{equation}
where $V_{\rm{obs}}$ is the observed rotation curve (see left panel Figure 5), $V_{\rm{Total}}$ is the total rotation curve (see Equation 1), and $V_{\rm{err}}$ the error bar in $V_{obs}$. The uncertainty on the observed rotation curve is discussed in \S 3.2. We use a differential evolution algorithm implemented in the python package Scipy \citep{virtanen2020scipy} for finding the maximum likelihood solution. 
We derive mass models by fixing the value of the stellar mass-to-light ratio derived using the stellar population synthesis models assuming a $'diet'-Salpeter$ initial mass function (IMF) (Table 6). The $'diet'-Salpeter$ function gives the maximum stellar mass in a photometric band. 

\subsection*{\underline{Dark Matter Halo}:} 
The results of the mass modelling are presented in Figure 10 and summarized in Table 7. From mass modelling, we find that the cuspy NFW halo ($\chi^{2}_{red} = 0.35$)  fits the observed rotation curve better than the cored PIS halo ($\chi^{2}_{red} = 2.4$). 
Mass models derived using PIS halo are characterized by a small core radius $R_c$ equal to 0.9 kpc, indicating that the baryons in  FGC 2366 are embedded in a compact dark matter halo defined as $R_{c}/R_{d} < 2$ \citep{banerjee2013some}. Mass models employing the NFW halo have a concentration parameter $c \sim 7.9$, higher than the median $c \sim 3.7$ of the ordinary disc galaxies of the THINGS galaxy sample \cite{de2008high}. In Figure 11, we compare the results obtained from the mass models of FGC 2366 with those of another extremely thin galaxy FGC 1440, and other superthin galaxies for which the mass models were already available in the literature (IC 2233, IC 5249, FGC 1540, UGC 7321). As a function of $a/b$, we plot (i) the ratio of the core radius of the PIS Dark Matter halo to the stellar disc scalelength $R_c/R_d$ (ii) $V_{\rm{rot}}/(R_{c}/R_{d})$ where $V_{\rm{rot}}$ is the asymptotic rotational velocity, and finally (iii) the concentration parameter $c$ from the NFW model with the respective regression lines superposed in each case. We note that, as expected, $R_{c}/R_{d}$ decreases with $a/b$, i.e., a thinner galaxy has a smaller value of $R_{c}/R_{d}$. \cite{banerjee2010dark} define a compact halo as one in which $R_{c}/R_{d} <2$.  Hence a compact dark matter halo could regulate the galactic disc dynamics from the inner galacto-centric radius. \cite{banerjee2013some} and  \cite{kurapati2018mass} argue that a compact dark matter halo is responsible for superthin stellar discs in these galaxies. We further observe that  $V_{\rm{rot}}/(R_{c}/R_{d})$ sharply increases with $a/b$. $V_{\rm{rot}}$ is a proxy of the total dynamical mass of the galaxy. In the case of superthin and extremely thin galaxies, the total dynamical mass is dominated by the dark matter halo. Thus $V_{\rm{rot}}/(R_{c}/R_{d})$ indicates the dark matter dominance in the inner galaxy, i.e., within the stellar disc. Finally, we find that the galaxies with a larger $a/b$ have larger values of $c$, indicating that thinner galaxies have higher concentration parameters in line with the findings of \cite{banerjee2010dark}.

\begin{figure}
\resizebox{80mm}{70mm}{\includegraphics{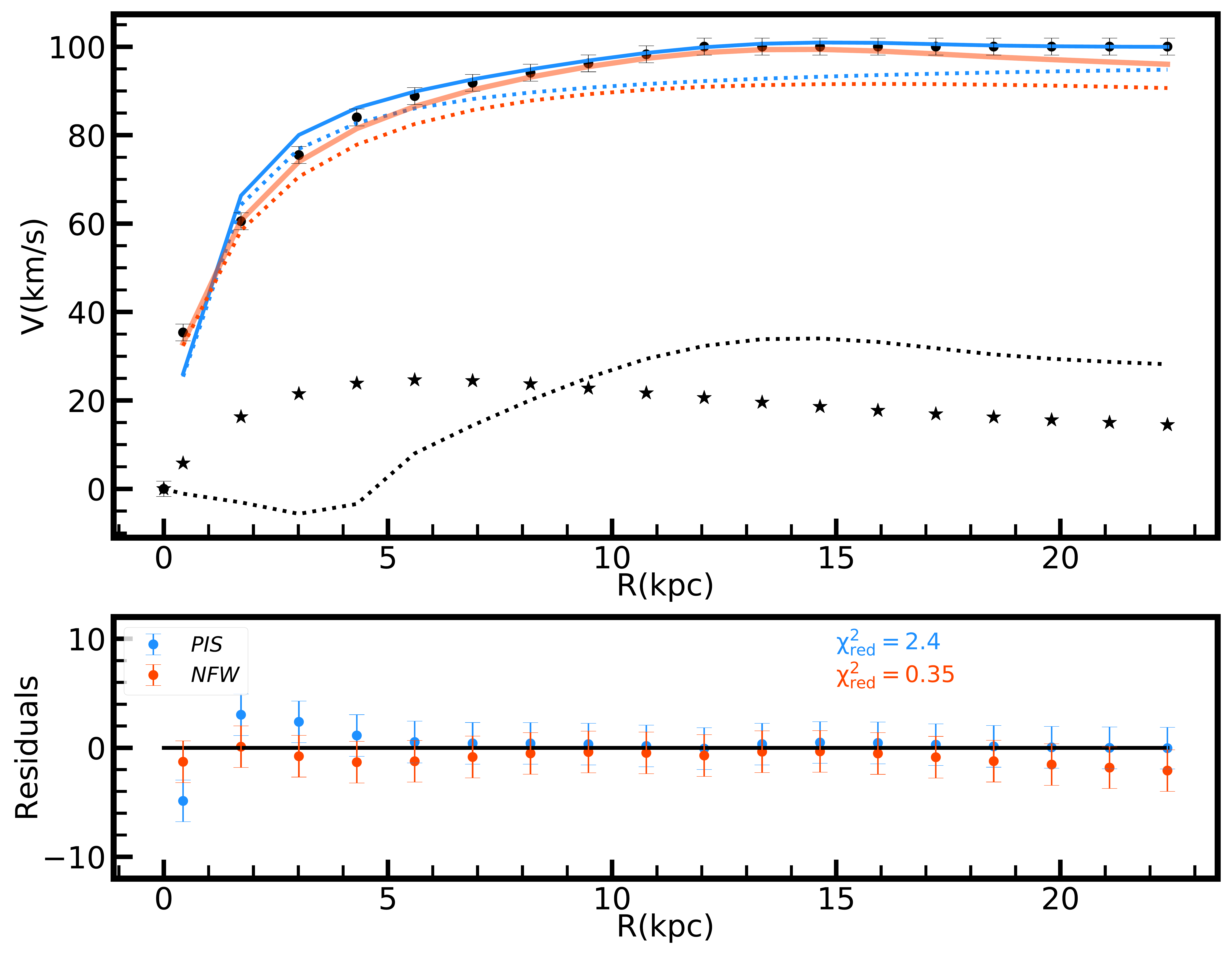}} 
\vspace{1pt}
\caption{In the top panel, we show the mass-models of the galaxy FGC 2366 derived using SDSS $z$-band photometry. The mass models are obtained by fixing $M/L$ from population synthesis models using a diet-Salpeter IMF. In the bottom panels, we show the residuals obtained by subtracting the data from the model weighed by the errors on the observed rotation curve. In the top panel, $stars$ and the $dashed$ curve indicate the circular velocities of stars and gas, respectively. The $dashed$ $blue$ and $red$ coloured curves indicate the circular velocity of the dark matter halo using a PIS and NFW models, respectively. Similarly, the $solid$ $blue$ and $red$ lines indicate the total rotation curve assuming a PIS and NFW dark matter model. The black points depict the observed rotation curve.}	
\end{figure}

\begin{table*}
\hspace{-65.5mm}
\begin{minipage}{110mm}
\hfill{}
\caption{Dark matter density profiles derived from mass-models}
\centering
\begin{tabular}{|l|c|c|c|c|c|c|c|c|}
\hline
\hline
Model                   &$c^{(a)}$ &$R^{(b)}_{200}$&$M/L^{(c)}$& $\chi^{2 (d)}_{red}$     & $\rho^{(e)}_{0}\times10^{-3}$  &$R^{(f)}_{c}$ & ${M/L}^{(g)}$& $\chi^{2 (h)}_{red}$  \\
                        &                           &(kpc)                           &                                &        &  $M_{\odot}/pc^{3}$                             &(kpc)                          &                                 &        \\
\hline
                        z-band                  & NFW  profile              &                                &                                &        &PIS profile                                      &                               &                                 &              \\
\hline

$'diet'$ Salpeter	&$ 7.9 \pm 0.2 $   &$ 58.3  \pm 0.4 $&$1.1$&$0.35$      		&$217.8\pm40.2$& $0.9  \pm  0.09$ &$1.1$&$2.4$\\

\hline
\end{tabular}  
\hfill{}
\label{table: table 7}
\end{minipage}
\begin{tablenotes}
\item  $(a)$: Concentration parameter of the NFW profile\\
\item $(b)$:  Radius at which the mean density equals 200 times the critical density.\\
\item $(c)$:  Mass to light ratio derived using population synthesis models or estimated as a free parameter. \\
\item $(d)$:  Reduced chi-square value corresponding to the fit. \\
\item $(e)$:  The central dark matter density of the PIS dark matter halo model\\
\item $(f)$:  The core radius of the PIS dark matter halo model\\
\item $(g)$:  Mass to light ratio derived using population synthesis models or estimated as a free parameter. \\
\item $(h)$:  Reduced chi-square value corresponding to the fit.\\
\end{tablenotes}
\end{table*}

\begin{figure*}
\vspace{-10.0pt}
\resizebox{180mm}{40mm}{\includegraphics{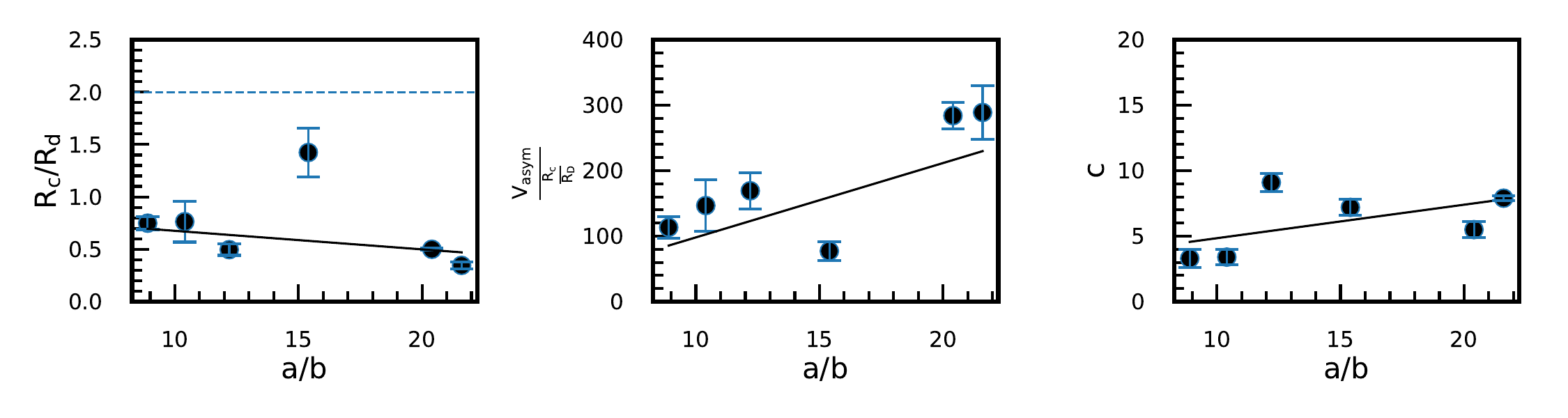}} 
\vspace{1pt}
\caption{For a sample of superthin and extremely thin galaxies including FGC 2366, we plot the ratio of the dark matter core radius to the disc scale length $R_c/R_d$ indicating its compactness in the left panel, the ratio of asymptotic rotation velocity $(V_{asym})$ to the compactness parameter $(R_{c}/R_{d})$ indicating the compactness of the mass distribution versus $a/b$ in the middle panel, and the dark matter concentration $c$, versus $a/b$ in third panel. In left panel, the horizontal the line indicates the limiting value of $R_{c}/R_{d}$ for a compact halo. Superimposed on the data points in each panel is the best-fitting regression line. The last two points in each panel represent the extremely thin galaxies FGC 1440 and FGC 2366, respectively.}
\end{figure*}

With the total mass inventory in place, we proceed to derive the dynamical models of FGC 2366 by using first the 2-component model of gravitationally-coupled stars + gas in the force field of the dark matter halo \textcolor{black}{(See \cite{10.1093/mnras/stab155}, for details)}. In the case of the 2-component model, we can not derive the radial velocity dispersion as the thin disc approximation, and several other simplifying assumptions decouple the vertical and radial motions. However, it can be used to derive the stellar vertical velocity dispersion using the stellar vertical scale height as a constraint. Next, we use this stellar vertical velocity dispersion as an initial input. We set up the distribution function-based equilibrium models using AGAMA \citep{vasiliev2019agama}  to derive the radial velocity dispersion and the stellar vertical scale height. If the latter matches the observed scaleheight, we conclude that the 2-component model and the Self-consistent model (SCM) are consistent.

\subsection{Self-consistent model of the stellar disc using AGAMA:}
We model the dynamics of FGC 2366 in dynamical equilibrium using the multi-component, self-consistent, iterative modelling method \citep{binney2014self, piffl2015bringing}  implemented in AGAMA \citep{vasiliev2018agama} in order to derive the vertical and the radial velocity dispersion of the stars. We model FGC 2366 as a 2-component system of the stars and the dark matter halo in the force field of the \HI{} gas. The stellar and the dark matter halo are defined by their distribution functions conforming to their corresponding density profiles; 
\HI{} is included as a static density component using the parameters from the tilted-ring models (\S 3.2). The details of the model profile are discussed below. We model the stellar disc with an exponential surface density consistent with the parameters obtained from the optical photometry. We model the stellar distribution function using the quasi-isothermal distribution function.

The stellar density profile is given by
\begin{equation}
 \Sigma_{s} (R) =\Sigma_{0,s}e^{\frac{-R}{R_{d,s}}}e^{\frac{-z}{h_{z,s}}}
\end{equation}
where, $\Sigma_{0,s}$ is the central stellar density, $R_{d,s}$ is the disc scalelength and $h_{z,s}$ is the exponential scaleheight. The quasi-isothermal distribution function for the stellar disc is given as follows:

\begin{equation}
 f(J)=f_{0}(J_{\phi}) \frac{\kappa}{\sigma_{R}^{2}} e^{- \frac{-\kappa J_{R}}{\sigma_{R}^{2}}}\frac{\nu}{\sigma_{z}^{2}}e^{\frac{-\nu J_{z}}{\sigma_{z}^{2}}}
\end{equation}
where $\kappa$ and $\nu$ are the radial and vertical epicyclic frequencies, respectively. $\sigma_{R}$ and $\sigma_{z}$ are the stellar velocity dispersion in
the respective directions. $J_{R}$, $J_{z}$, and $J_{\phi}$ are the actions of the stellar discs in the corresponding directions. 
Here, $J^{2}_{\phi}=R^{3} \frac{\partial \Phi }{\partial R}$, $J^2 = {J_R}^2 + {J_{\phi}}^2 + {J_{z}}^2$ and  
$f_{0}(J_{\phi}) = \frac{\Sigma(R) \Omega (R)}{2 \pi^{2} \kappa^{2}(R) }$, where $\Phi$ and $\Omega$ are the total gravitational potential and angular velocity 
respectively.  The radial dependence of the velocity dispersion in the radial and vertical directions is modelled as

\begin{equation}
\!
\begin{aligned}
\sigma_{R}(R)&= \sigma_{R,0}e^{\frac{-R}{R_{\sigma_{R}}}}\\
\sigma_{z}(R)&= \sigma_{z,0}e^{\frac{-R}{R_{\sigma_{z}}}} 
\end{aligned}
\end{equation}
 
where $R_{\sigma_{R}}$ and $R_{\sigma_{z}}$ are the scale lengths in the radial and the vertical directions, respectively.

The \HI{} density is modelled as a static potential as follows:

\begin{equation}
  \Sigma_{g} (R) =\Sigma_{0,g}e^{-(\frac{R}{a_{g}})^{2}}
\end{equation}
where the values of $\Sigma_{0,g}$ and $a_{g}$ are obtained by fitting the \HI{} surface density derived from our \HI{} observations using the tilted
ring modelling (see Figure 5). 

The dark matter, modelled as a 3-parameter spherically-symmetric function, is given by:

\begin{equation}
 \rho_{r}= \rho_{0} (r/a)^{-\gamma}\bigg(1+\bigg(\frac{r}{a}\bigg)^{\alpha} \bigg)^{(\frac{\gamma - \beta}{\alpha})}
\end{equation}

We use $\alpha=1$, $ \beta=3$, $\gamma =1$, in the above to mimic the cuspy NFW profile, where $\rho_{0}$ is the central density and $a$ is the characteristic scale of the NFW halo given by $a=R_{200}/c$. Similarly, if we set $\alpha=2$, $ \beta=2$, $\gamma =0$, for the cored PIS halo, $a$ mimics the core radius of the density profile and $\rho_{0}$ the value central density. We derive the distribution function for the dark matter halo corresponding to the input density profile by the \textit{Quasi-Spherical} distribution instance implemented in AGAMA. It constructs the distribution function using the input density profile using the generalized Eddington inversion formula. It then uses a spherical action finder to convert the distribution function into an action-based form. The values
for the dark matter density for the PIS and NFW halo are taken from the mass model derived in \S 5 (Table 7). We use the value of the stellar vertical velocity dispersion obtained from the 2-component method (See \cite{10.1093/mnras/stab155}) as input for constructing the self-consistent models using AGAMA. We start by creating an initial guess for the total potential of FGC 2366 using the results obtained from tilted ring modeling, optical photometry, and mass models. We then construct actions corresponding to this potential using the "action-finder" implemented in AGAMA and compute the distribution function and the density from the distribution function. We again calculate the updated potential by solving Poisson's equation for the combined density of all the terms. We then repeat the above procedure of calculating the actions from the updated potential and evaluate the distribution function at every step. We compute the density from the distribution function and calculate the potential until the potential converges.

\subsection*{\underline{Stellar velocity dispersions}} 
In Figure 12, we present the best-fitting SCM for FGC 2366 obtained using the NFW dark matter profile, which has a smaller value of the reduced ${\chi}^2$ than the PIS model. We note that the central vertical stellar velocity dispersion $\sigma_{z,0}=23$ \kms is comparable to that of the Milky Way thin disc ($\sigma_{z,0}=25$  \kms) \citep{sharma2014kinematic}. 

Interestingly, the value of  $\sigma_{R,0}=61 $\kms for FGC 2366 is higher than the value of $\sigma_{R,0}=40 \kms$ obtained for Milky Way. \citep{sharma2014kinematic, bovy2012milky}. In Figure 13, we present $\rm{Min}(\sigma_{z}/\sigma_{R})$  as a function of $a/b$ for our galaxy sample, overlaying the corresponding regression line. A small value of $\sigma_{z}/\sigma_{R}$ indicates that the excursion of the stars from their mean vertical position in the galactic midplane is smaller than their dispersion about a mean radius in the plane. For field galaxies like the superthin, an increase in $\sigma_{z}$ is a reflection of disc thickening due to vertical bending instabilities and counter-rotating bars among others \citep{khoperskov2003minimum,khoperskov2017disk}; a large value of $\sigma_{R}$, on the other hand, is attributed to disc heating by bars and spiral arms, and also to radial migration of stars \textcolor{black}{\citep{rovskar2013effects}}. \cite{jenkins1990spiral} shows that the relatively higher rate of dynamical heating in the plane renders the value of $\sigma_{z}/\sigma_{R}$ smaller. Due to the edge-on orientation of these galaxies, it is not possible to ascertain the presence of the spiral structure in their plane. Thus it leaves the question regarding the dynamical origin of the high anisotropy of the stellar velocity ellipsoid in these galaxies. 

\begin{figure*}
\resizebox{180mm}{45mm}{\includegraphics{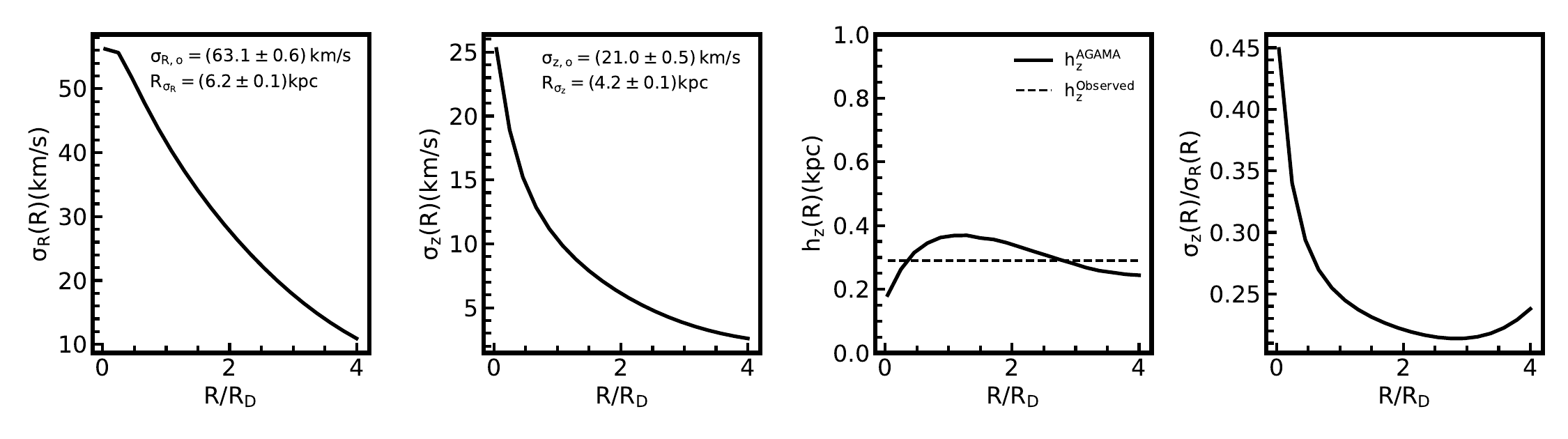}} 
\caption{In first panel, we show the stellar radial velocity dispersion, 
in panel 2, we plot the stellar vertical velocity dispersion, in the third panel, we show the stellar scaleheight, and in the fourth panel we show the ratio of vertical-to-radial stellar velocity dispersion, as a function of galactocentric radius normalized by disc scale length, as determined by dynamical modelling with an NFW dark matter halo using AGAMA for FGC 2366. The dotted line in third panel shows the observed scale height.}
\end{figure*}

\begin{figure}
\resizebox{75mm}{60mm}{\includegraphics{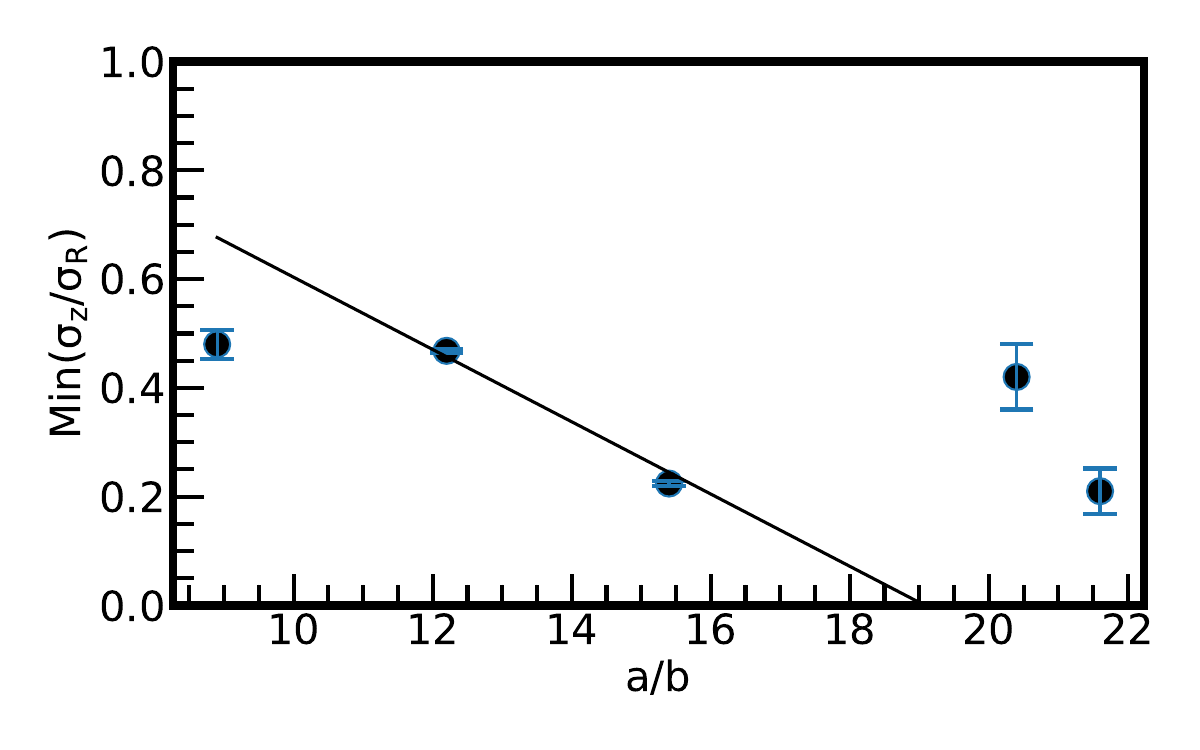}} 
\caption{Ratio of the vertical-to-radial stellar velocity dispersion as a function of the major-to-minor axes ratio derived from dynamical models for a sample of superthin and extremely thin galaxies, including FGC 2366. The regression line is superposed. The last two points in the plot represent the extremely thin galaxies FGC 1440 and FGC 2366, respectively.}	
\end{figure}

\subsection*{\underline{Disc dynamical stability of FGC 2366}}
 Since a hotter disc is dynamically stabler, the $\sigma_{R}$ value obtained above partially regulates the dynamical stability of the galactic discs, and therefore possibly their razor-thin morphology. We use the multi-component stability parameter proposed by \cite{romeo2011effective}:
\begin{equation}
\frac{1}{Q_{RW}} = \left\{
                \begin{array}{ll} \frac{W_{\sigma}}{T_{s}Q_{s}} + \frac{1}{T_{g}Q_{g}}  \hspace*{0.5cm} if \hspace*{0.5cm}  T_{s}Q_{s} > T_{g}Q_{g}\\  
         \frac{1}{T_{s}Q_{s}} +\frac{W_{\sigma}}{T_{s}Q_{s}}                            \hspace*{0.5cm} if \hspace*{0.5cm} T_{s}Q_{s} < T_{g}Q_{g}  
         \end{array}
              \right.
\end{equation}

\noindent where the weight function $\mathbf{W_{\sigma}}$ is given by
\begin{equation}
 \mathbf{W_{\sigma} =\frac{2\sigma_{s} \sigma_{g}}{\sigma_{s} ^{2} + \sigma_{g}^ {2}}}
\end{equation}
,the thickness correction defined as
\begin{equation}
 T \approx 0.8 + 0.7 \frac{\sigma_{z}}{\sigma_{R}}
\end{equation}

\noindent and the Toomre $Q$ parameters of the stellar and the gaseous discs are given by $Q= \frac{\kappa \sigma  }{\pi G \Sigma}$, where $\kappa$ is the epicyclic frequency, and $\Sigma$ the surface density. The $\sigma_{R}$, $\sigma_{z}$ and hence $\sigma_{s}$ values were estimated from self-consistent iterative modelling in section 5.2, and $\sigma_{g}$ from the tilted ring modelling from section 3.2. A value of $Q_{RW} >1$ should indicate that the disc is stable against the growth of local axisymmetric perturbations. However,
numerical studies have shown that in non-axisymmetric perturbations, a $Q_{RW} \sim 2-3$ indicates a stable disc. We have presented the $Q_{RW}$ as a function of $R$ for FGC 2366 in Figure 14. The minimum value of $Q_{RW}=3.1$ indicates that the disc is very stable against the growth of axisymmetric instabilities. In Figure 15, we compare the $Q_{RW}$ values at 1.5$R_{d}$ of the extremely thin galaxy FGC 2366 with a previously studied sample of superthin galaxies and with another extremely thin galaxy FGC 1440. The corresponding regression line in Figure 16 indicates that the values of $Q_{RW}$ at 1.5$R_{D}$  increase with $a/b$, indicating that the galaxies with the thinnest discs have the highest disc dynamical stability values; see also, \cite{garg2017origin}.

\begin{figure}
\resizebox{75mm}{60mm}{\includegraphics{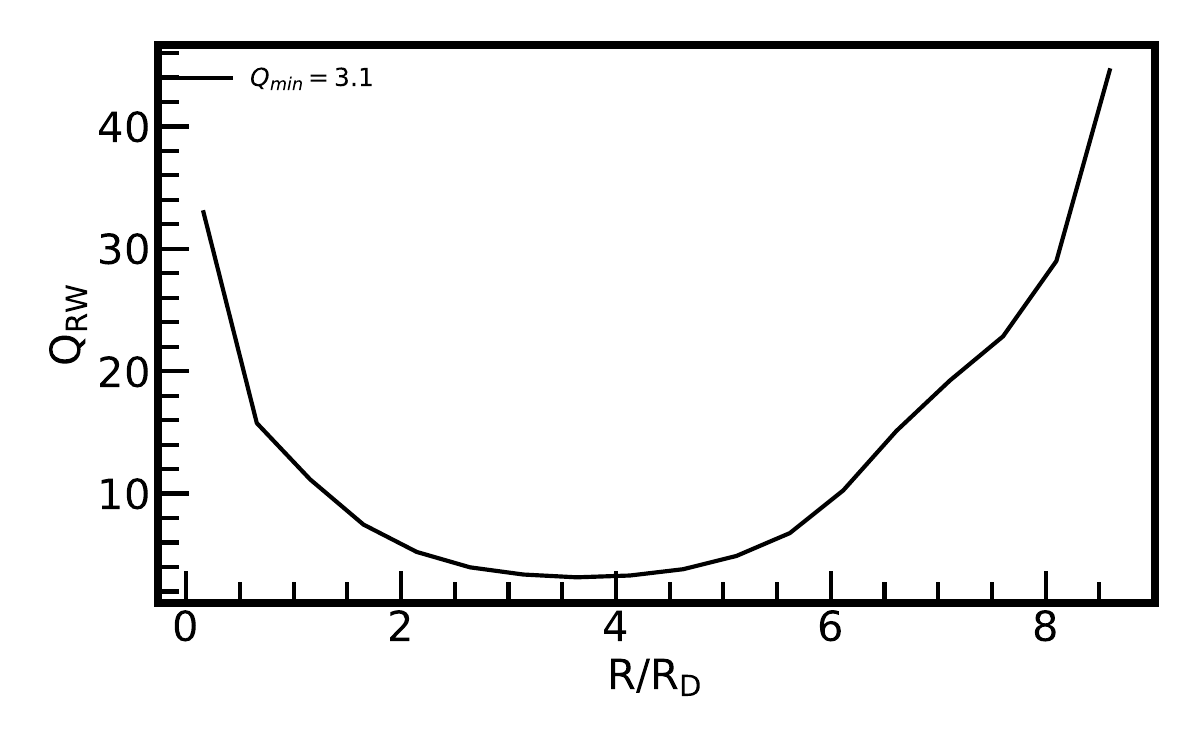}} 
\caption{ 2-component disc dynamical stability parameter $Q_{RW}$ as a function of $R/R_D$ for FGC 2366 as obtained from AGAMA.}
\end{figure}

\begin{figure}
\resizebox{75mm}{60mm}{\includegraphics{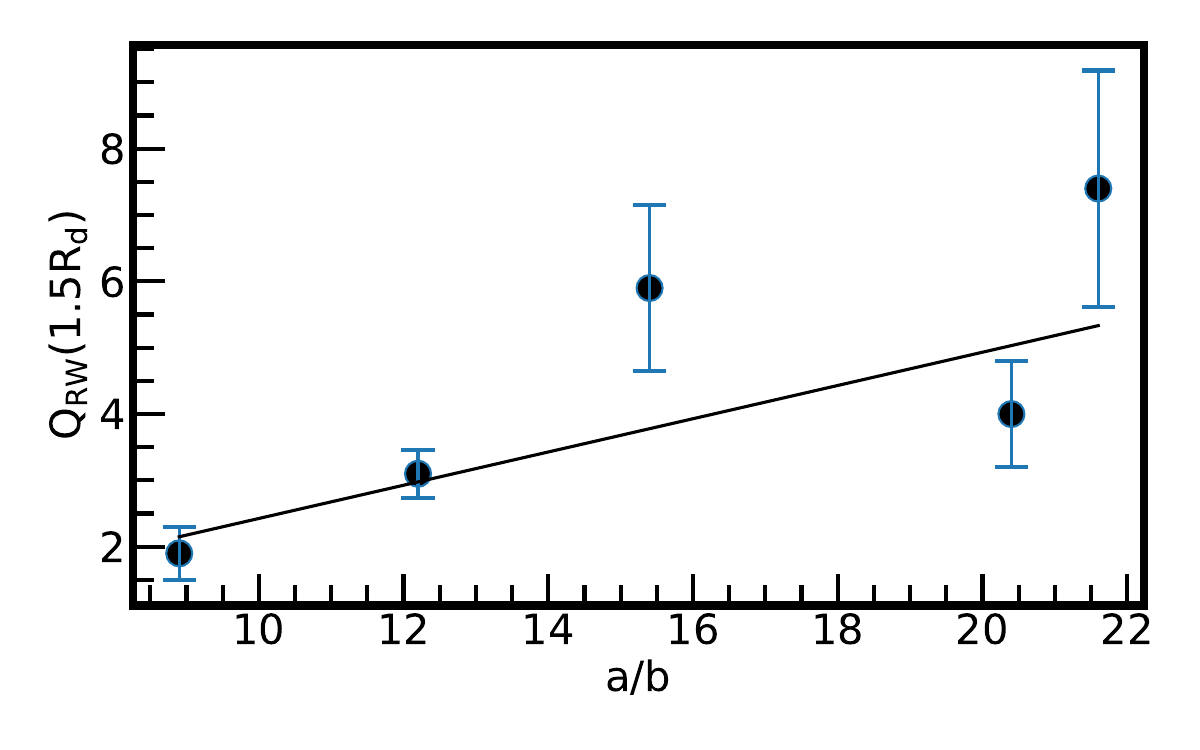}} 
\caption{2-component disc dynamical stability parameter $Q_{RW}$ at $R=1.5 R_d$ as a function of the major-to-minor axes ratio $a/b$ for a sample of superthin and extremely thin galaxies including FGC 2366. The last two points represent the extremely thin galaxies FGC 1440 and FGC 2366, respectively.}
\end{figure}

\subsection*{\underline{Specific angular momentum}}
The specific angular momentum $j_{i}$ of the $i^{th}$ component of the disc may be given by

\begin{equation}
 j_{i}(R)= \frac{2 \pi \int^{R} _{0} R^{'2} \Sigma_{i}(R^{'}) V(R^{'})dR^{'} }{ 2 \pi \int^{R} _{0} R^{'} \Sigma_{i}(R^{'}) dR^{'}}
\end{equation}

\noindent where $\Sigma_{i}(R)^{'}$ corresponds to the surface density of the $i^{th}$ component, and 
$V$ the asymptotic rotational velocity. In Figure 16, we present the  j*- M relation for ordinary bulgeless disc galaxies as derived by \citet{jadhav2019specific}, and also that of a sample of disc galaxies spanning a large mass range  $7\leq \rm{log}(M_{*}/M_{\odot}) \leq11.5$ and 
$6\leq \rm{log}(M_{gas}/M_{\odot}) \leq11$ as obtained by \cite{2021A&A...647A..76M}. We superpose on them the data points for FGC 2366 along with another extremely thin galaxy, FGC 1440 studied earlier, and also those of the sample of superthins from \citet{jadhav2019specific}. Unlike FGC 1440, we note that FGC 2366 lies above the 95.4$\%$ confidence intervals of both the $j{*}-M_{*}$ relations, similar to the trend in other superthin galaxies. Figure 17, we plot $j_{*}$ as a function of $a/b$ for the sample of superthins studied earlier along with FGC 1440 and FGC 2366. The regression line fit indicates that $j_{*}$ increases with $a/b$, meaning thinner discs may have higher specific angular momentum.

\subsection*{\underline{Principal Component Analysis}}

Finally, we attempt to understand the physical factor primarily driving the vertical structure of the stellar discs in superthin galaxies, the possible parameters being $j_{*}$, $Q_{RW}(1.5 R_d)$, 
$\rm{Min {(\sigma}_z/{\sigma}_R)}$, for the stars $V_{\rm{rot}}/(R_c/R_d)$  along with $a/b$. In Table 8, we present the same for FGC 2366 as calculated from the dynamical models obtained in this paper. The same can be obtained for all the superthins studied earlier and the extremely thin galaxy FGC 1440. Since we have less than ten galaxies in our sample (IC 2233, FGC 1540, UGC 7321, FGC 1440, \& FGC 2366), we create $\sim$ 1000 mock galaxies from each galaxy of our sample as follows. For each galaxy, we construct a 5-D Gaussian distribution of the parameters using the best-fitting value of each parameter as the mean and the error as the standard deviation for the corresponding distribution. We then randomly sample 1000 points from this 5-D parameter space which serve as the mock galaxies. We repeat this process for each galaxy in our sample. We next carry out a Principal Component Analysis (PCA) of the parameters: $j_{*}$, $Q_{RW}(1.5 R_d)$, $\rm{Min}{(\sigma}_z/{\sigma}_R)$, $V_{\rm{rot}}/(R_c/R_d)$  along with $a/b$. The principal components (PCs) are presented in Figure 18. In Figure 19, we present the decomposition of the PCs in terms of the different parameters. We note that the contributions from $a/b$, $V_{\rm{rot}}/Rc/RD$ and $Q_{RW}$ is the maximum for PC1 and PC2 taken together and explains 88$\%$ of the variation in data. This possibly indicates that the major driving factor for the superthin vertical structure is the dark matter dominance at inner galactocentric radii as indicated by $V_{\rm{rot}}/(R_{c}/R_{d})$.

\begin{table}
\hfill{}
\caption{Dynamical parameters of FGC 2366 as obtained using AGAMA}
\centering
\begin{tabular}{|l|c|}
\hline
\hline
Parameter                                       & Value         \\
\hline
$R_{c}/R_{d}$                                   &     $0.35 \pm 0.03$                \\ 
$V_{\rm{rot}}/(R_{c}/R_{d})$                    &     $288  \pm 40 $          \\
$\sigma_{z}/\sigma_{R}$                         &     $0.2 \pm  0.04 $             \\
$Q_{RW}(1.5R_{d})$                              &     $7.0  \pm 1.8$               \\
$\rm{log}_{10}( j_{*}/ \rm{kpc} \kms{}) $       &     $2.67 \pm 0.02 $             \\ 
\hline   
\end{tabular}
\hfill{}
\label{table: table 8}
\end{table}

\begin{figure}
\resizebox{75mm}{60mm}{\includegraphics{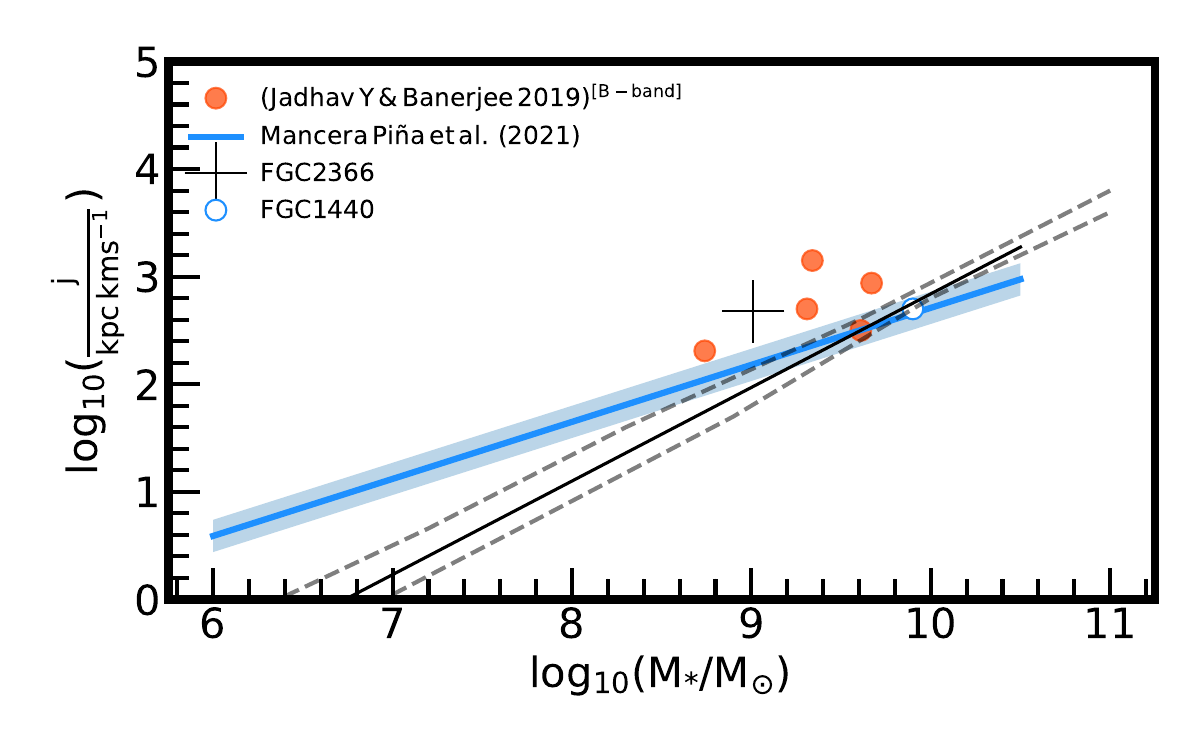}} 
\caption{Comparison of the specific angular momentum of stars in FGC 2366 with a sample of previously studied superthin galaxies and the extremely thin galaxy FGC 1440. The $'blue'$ regression line is for a sample of disc galaxies studied by \protect\cite{2021A&A...647A..76M}. The solid black line shows the regression line obtained by \protect\cite{jadhav2019specific} for a sample of six ordinary bulgeless disc galaxies with bulge fraction less than 0.05 from a larger sample of disc galaxies studied by \protect\cite{obreschkow2014fundamental}. The dashed line shows the 95.4 $\%$ confidence interval on the regression line obtained by \protect\cite{jadhav2019specific}.}
\end{figure}

\begin{figure}
\resizebox{75mm}{60mm}{\includegraphics{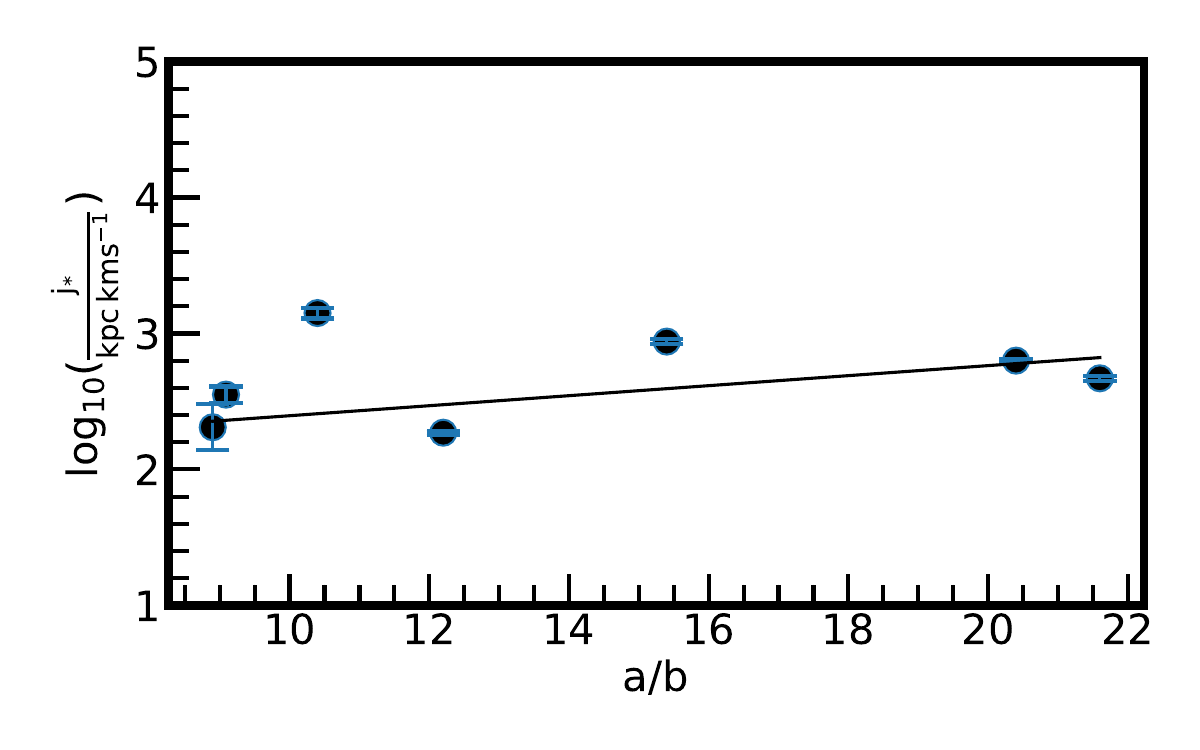}} 
\caption{Stellar specific angular momenta $j*$ as a function of the major-to-minor axes ratio $a/b$ with a sample of previously studied superthin galaxies and the extremely thin galaxy FGC 1440. The last two points represent the extremely thin galaxies FGC 1440 and FGC 2366, respectively.}
\end{figure}

\begin{figure}
\resizebox{75mm}{60mm}{\includegraphics{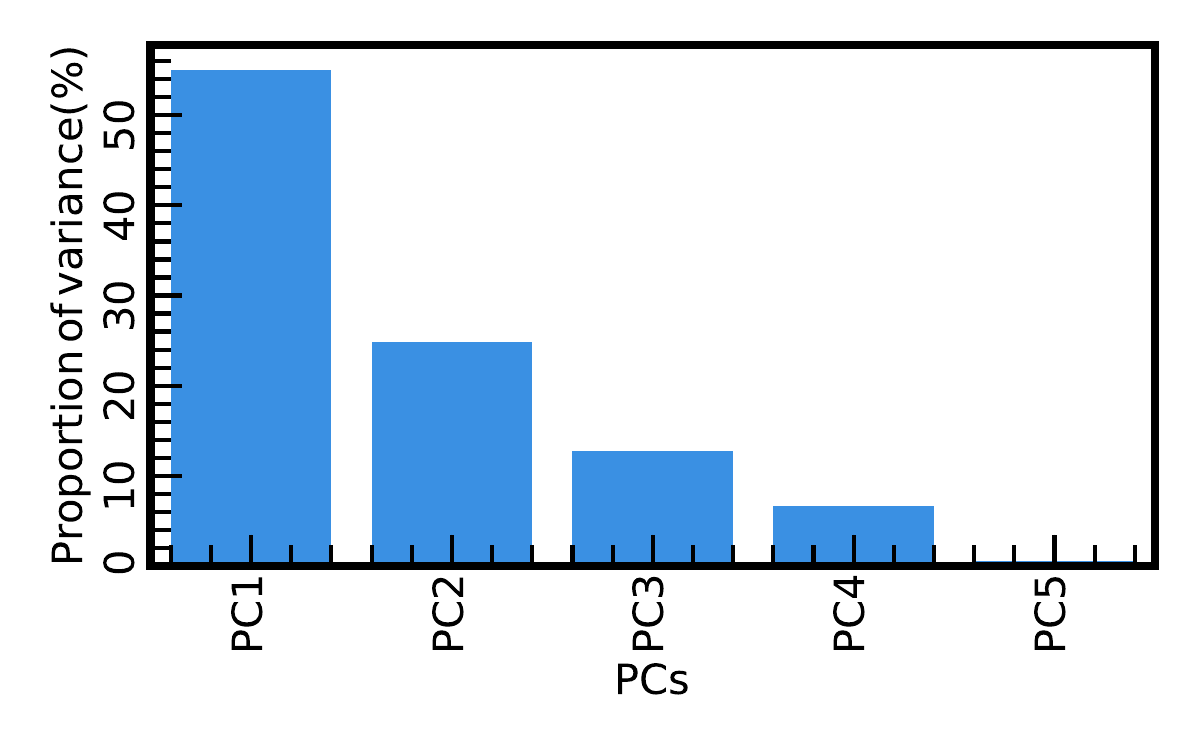}} 
\caption{Scree plot from the principal component analysis of the dynamical parameters plausibly responsible for driving the superthin stellar discs.}	
\end{figure}

\begin{figure}
\resizebox{100mm}{85mm}{\includegraphics{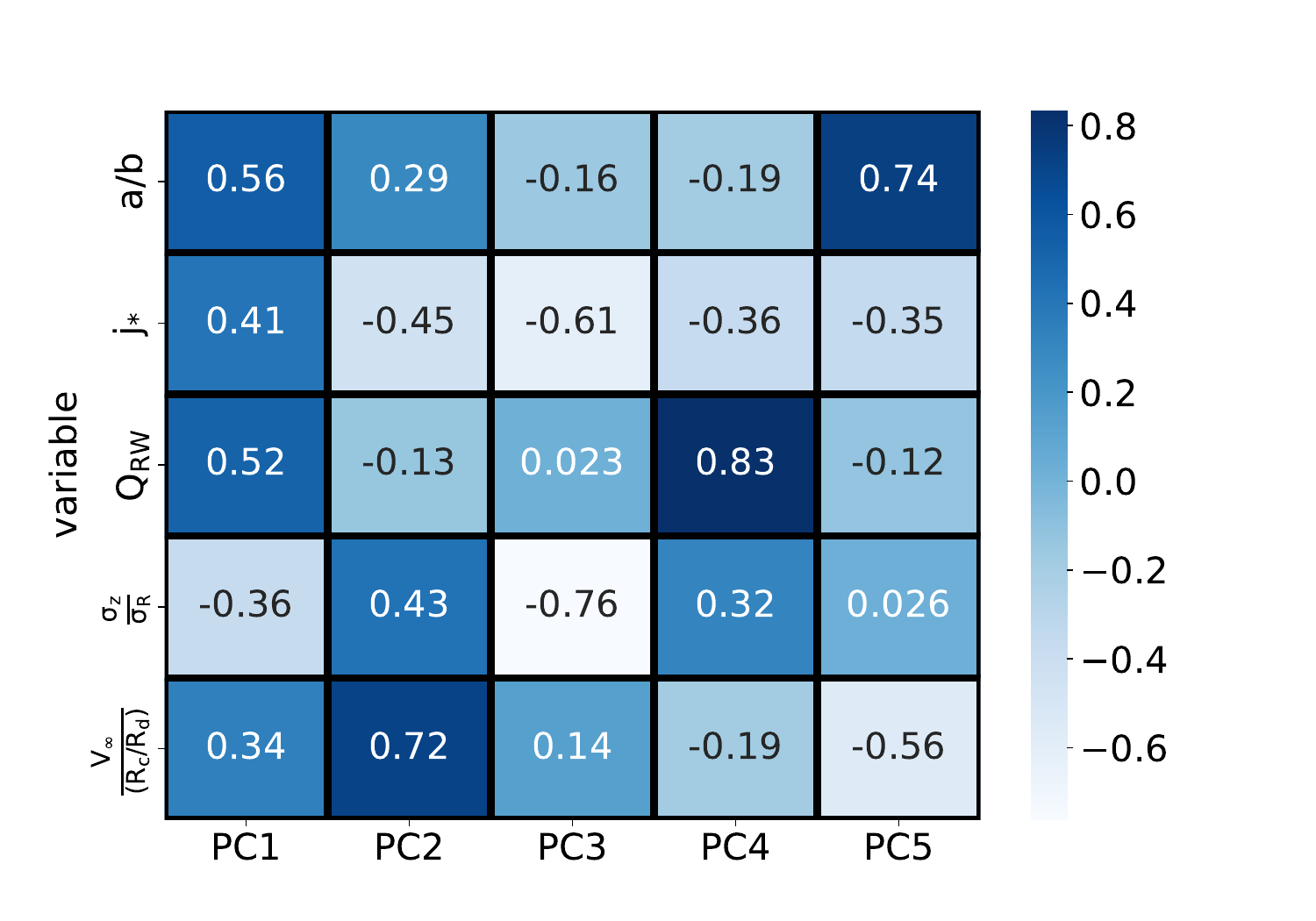}} 
\caption{Loadings of the Principal Component Analysis of the dynamical parameters plausibly responsible for driving the superthin stellar discs. }	
\end{figure}

\section{Conclusions}

In this paper, we present GMRT \HI{} 21cm radio-synthesis observations of an edge-on low surface brightness galaxy FGC 2366 with the largest axial ratio a/b = 21.6 and hence the thinnest stellar disc. We carry out the data reduction using CASA and employ the 3-D tilted ring modelling to derive the kinematics and structural properties of neutral \HI{} distribution. \emph{We note that \HI{} distribution is uniform and unperturbed and does not show a signature of warping or flaring in FGC 2366.} \\

\noindent We also present the optical photometry in the $g$, $r$ and $z$ bands. Using the total \HI{} rotation curve along with the $z$-band optical photometry, we next derive its mass model, and, after that, we construct the dynamical model of the stellar disc and the dark matter halo using the distribution-function-based stellar dynamical code AGAMA using a quasi-isothermal distribution function to represent the stars and a quasi-Spherical one for dark matter halo. \emph {We find that the NFW dark matter halo gives a better fit to the steeply rising rotation curve of FGC 2366.}\\

\noindent Finally, we consider a set of parameters obtained from the dynamical model constructed that may be responsible for the superthin vertical structure of the stellar discs: the specific angular momentum of the stellar disc $j_{*}$, the 2-component disc stability parameter at $R=1.5 R_d$ $Q_{RW}(1.5 R_d)$, the ratio of the central values of the stellar vertical velocity dispersion -to-the-radial velocity dispersion $\rm{Min}{(\sigma}_z/{\sigma}_R)$, and the compactness of the mass distribution $V_{\rm{rot}}/(R_{c}/R_{d})$. In order to identify the dynamical parameter/s crucial for the existence of the superthin vertical structure, we carry out a Principal Component Analysis (PCA) of the above parameters corresponding to a sample of superthin galaxies and the extremely thin galaxy FGC 1440 studied in the past along with those of FGC 2366. We found that the variance in the values of the above parameters was mostly explained by $V_{\rm{rot}}/(R_{c}/R_{d})$. \emph{We, therefore, conclude that the compactness of the mass distribution is fundamentally responsible for a superthin stellar vertical structure.}
 
\section{Data Availability}
The data from this study are available upon request.

\section{ACKNOWLEDGEMENT}
We thank the staff of the GMRT that made these observations possible. GMRT is run by the National Centre for Radio Astrophysics of the Tata Institute of Fundamental Research. 
Aditya would like to thank Dr. Eugene Vasiliev for help with AGAMA. PK is partially supported by the BMBF project 05A20PC4 for D-MeerKAT. SB and DM acknowledge the support by the Russian Science Foundation, grant 19-12-00145. We acknowledge the usage of the HyperLeda database. (\textcolor{blue}{http://leda.univ-lyon1.fr}). We also thank Dr. K. Saikranthi for the helpful discussion.

\small{\bibliographystyle{mnras}}
\bibliography{2366_d2}

\section{Appendix - A}

\begin{figure}
\hspace{-8mm}
\resizebox{85mm}{65mm}{\includegraphics{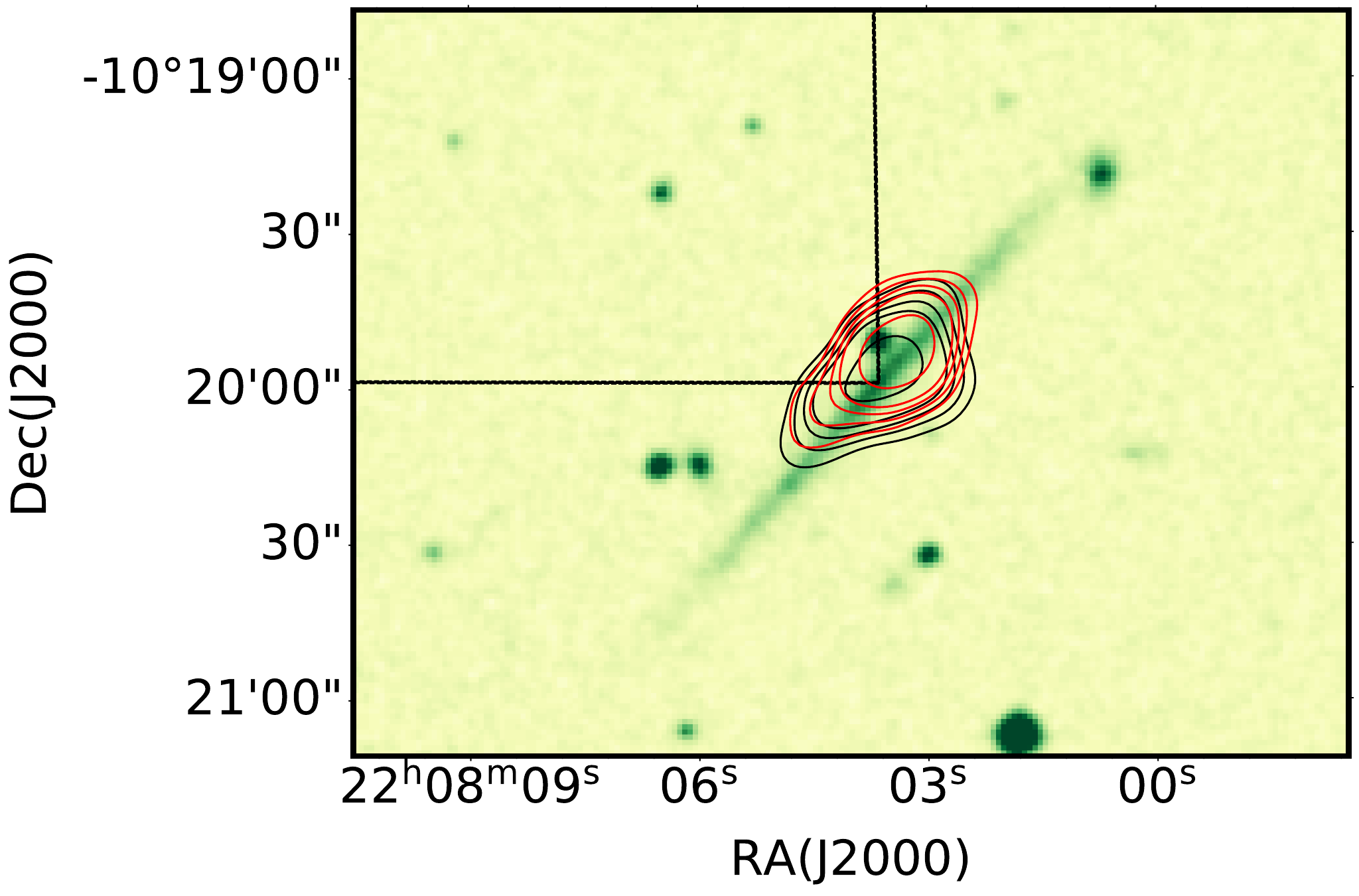}} 
\caption{Channel maps showing \HI{} emission from FGC 2366. The black contours show the emissions observed at 2832.5 \kms{}, which is systemic velocity obtained by fitting the 3D model to the observed data cube. The red contours, on the other hand, 
represent the emissions at 2844 \kms{} which is obtained by fitting the busy function to the 1D spectrum derived from the data cube.
The emissions at 2832.5 \kms{} coincide with the centre of the optical image than the emissions at 2844 \kms{}.
The contour levels are at [3, 4, 5,6, 9]$\times$ 0.8 mJy beam$^{-1}$}	
\end{figure}

\end{document}